\documentclass[journal,comsoc]{IEEEtran}
\usepackage[T1]{fontenc}

\usepackage{amssymb,amsmath,amsfonts,amsthm}
\usepackage{mathrsfs}
\usepackage{cite}
\usepackage{array}
\usepackage{tabularx}
\usepackage[table]{xcolor}
\usepackage{color}
\usepackage{mathtools}
\usepackage{subcaption}
\usepackage{caption}
\usepackage{mathtools}
\usepackage{graphicx}
\usepackage{bbm}
\usepackage[utf8]{inputenc}

\newtheorem{theorem}{Theorem}
\newtheorem{lemma}{Lemma}

\newcommand{\antdensity}{\mu}

\IEEEoverridecommandlockouts
\ifCLASSINFOpdf
\else
\fi
\usepackage[cmintegrals]{newtxmath}
\begin{document}
	\title{Channel Hardening and Favorable Propagation\\ in Cell-Free Massive MIMO with \\Stochastic Geometry}
	
	\author{Zheng~Chen,~\IEEEmembership{Member,~IEEE}, Emil~Bj\"{o}rnson,~\IEEEmembership{Senior Member,~IEEE}
		\thanks{Z. Chen and E. Bj\"{o}rnson are with the Department of Electrical Engineering (ISY), Link\"{o}ping University, Link\"{o}ping, Sweden (email: zheng.chen@liu.se, emil.bjornson@liu.se). A part of this work was presented at GLOBECOM Workshops 2017 \cite{conference-version}.}
		\thanks{This work was supported in part by ELLIIT, CENIIT, and the Swedish Foundation for Strategic Research (SSF).}}
\maketitle

\begin{abstract}	
Cell-Free (CF) Massive MIMO is an alternative topology for future wireless networks, where a large number of single-antenna access points (APs) are distributed over the coverage area. There are no cells but all users are jointly served by the APs using network MIMO methods.
Prior works have claimed that CF Massive MIMO inherits the basic properties of cellular Massive MIMO, namely channel hardening and favorable propagation. In this paper, we evaluate if one can rely on these properties when having a realistic stochastic AP deployment. Our results show that channel hardening only appears in special cases, for example, when the pathloss exponent is small.
 However, by using 5--10 antennas per AP, instead of one, we can substantially improve the hardening. Only spatially well-separated users will exhibit favorable propagation, but when adding more antennas and/or reducing the pathloss exponent, it becomes more likely for favorable propagation to occur.
The conclusion is that we cannot rely on channel hardening and favorable propagation when analyzing and designing CF Massive MIMO networks, but we need to use achievable rate expressions and resource allocation schemes that work well also in the absence of these properties. Some options are reviewed in this paper.
\end{abstract}

\section{Introduction}

The throughput of conventional cellular networks is limited by the uncoordinated inter-cell interference. To mitigate this interference, Shamai and Zaidel introduced the co-processing concept in 2001 \cite{Shamai2001a}, which is more commonly known as network multiple-input multiple-output (MIMO) \cite{Venkatesan2007a}. The key idea is to let all the access points (APs) in the network jointly serve all users, in downlink as well as uplink, thereby turning interference into useful signals \cite{Zhou2003a,Gesbert2010a}. Despite the great theoretical potential, network MIMO is challenging to implement and the 3GPP LTE standardization of the technology failed to provide any remarkable gains \cite{Fantini2016a}.

Two practical issues with network MIMO are to achieve scalable channel acquisition and sharing of data between APs. The former can be solved by utilizing only local channel state information (CSI) at each AP \cite{Bjornson2010c}, which refers to knowledge of the channels between the AP and the users. These channels can be estimated by exploiting uplink pilot transmission and channel reciprocity in time-division duplex (TDD) systems, thus making TDD a key enabler for network MIMO. User-centric clustering, where all APs reasonably close to a user transmit signals to it, is key to reducing the data sharing overhead \cite{Bjornson2011a}. These concepts are not supported by LTE, which is instead designed for codebook-based channel acquisition and network-centric AP clustering.

The network MIMO concept has recently reappeared under the name Cell-Free (CF) Massive MIMO \cite{Ngo2015a,Nayebi2015a}. The new terminology is used for networks consisting of massive numbers of geographically distributed single-antenna APs, which are used to jointly serve a set of users that is substantially smaller than the number of APs.
In particular, it has been presented as a better option for providing coverage than using uncoordinated small cells \cite{CF_mimo}. The CF concept is fundamentally the same as in the network MIMO paper \cite{Bjornson2010c}, where the APs perform joint transmission with access to data to every user but only local CSI. The main novelty introduced by CF Massive MIMO is the capacity analysis that takes practical pilot allocation and imperfect CSI into account \cite{CF_mimo,Nayebi2017a,Bashar2018a}, using similar methodology as in the cellular Massive MIMO literature~\cite{Marzetta2016a}. Although the initial releases of 5G are cell-based, the standard separates the control and user planes, thus cell-free data transmission is theoretically feasible, if all the practical issues are resolved \cite{Interdonato2019a}.

Each AP in a cellular Massive MIMO systems serves its own exclusive sets of users using an array with a massive number of co-located antennas; see Fig.~\ref{fig:system-model}(a). Such systems deliver high spectral efficiency by utilizing the \textit{channel hardening} and \textit{favorable propagation} phenomena \cite{Marzetta2016a}. Channel hardening means that the beamforming transforms the fading multi-antenna channel into an almost deterministic scalar channel \cite{channel_hardening}. This property simplifies the resource allocation, since there is no need to adapt the power allocation or scheduling to the small-scale fading variations \cite{SIG-093}. Favorable propagation means that the users' channel vectors are almost orthogonal \cite{favorable_propagation}. This is favorable since there will be little interference leakage between the users. Both phenomena are consequences of the law of large numbers.

\subsection{Limitations of Existing Work}

\begin{figure}[ht!]
\begin{subfigure}[b]{0.5\textwidth}
	\centering
	\includegraphics[width=0.8\textwidth]{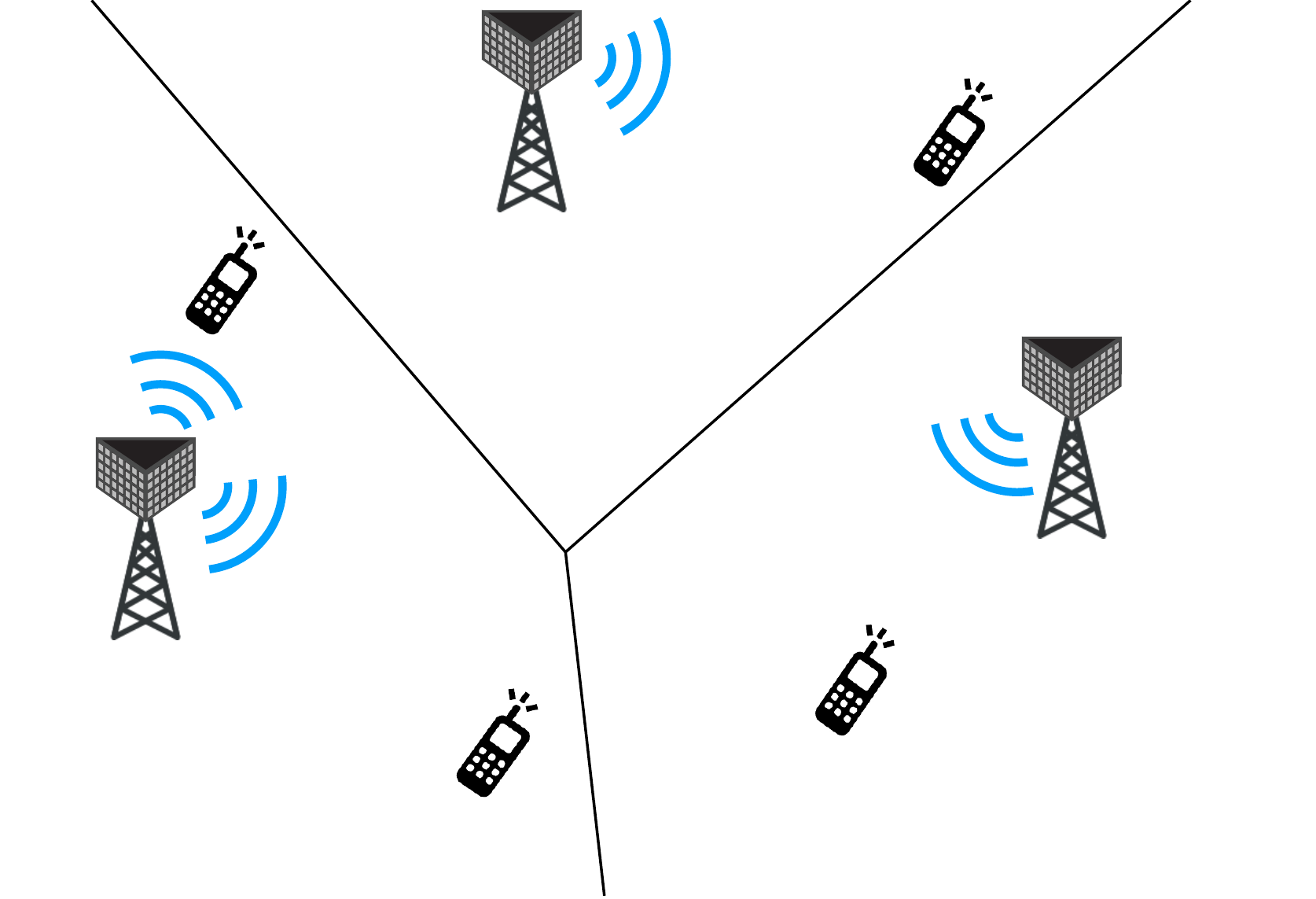}
	\caption{In a cellular Massive MIMO system, there are (relatively) few APs and each one is serving an exclusive subset of the users, using a large number of antennas.}
\end{subfigure}
\begin{subfigure}[b]{0.5\textwidth}
	\centering
	\includegraphics[width=\textwidth]{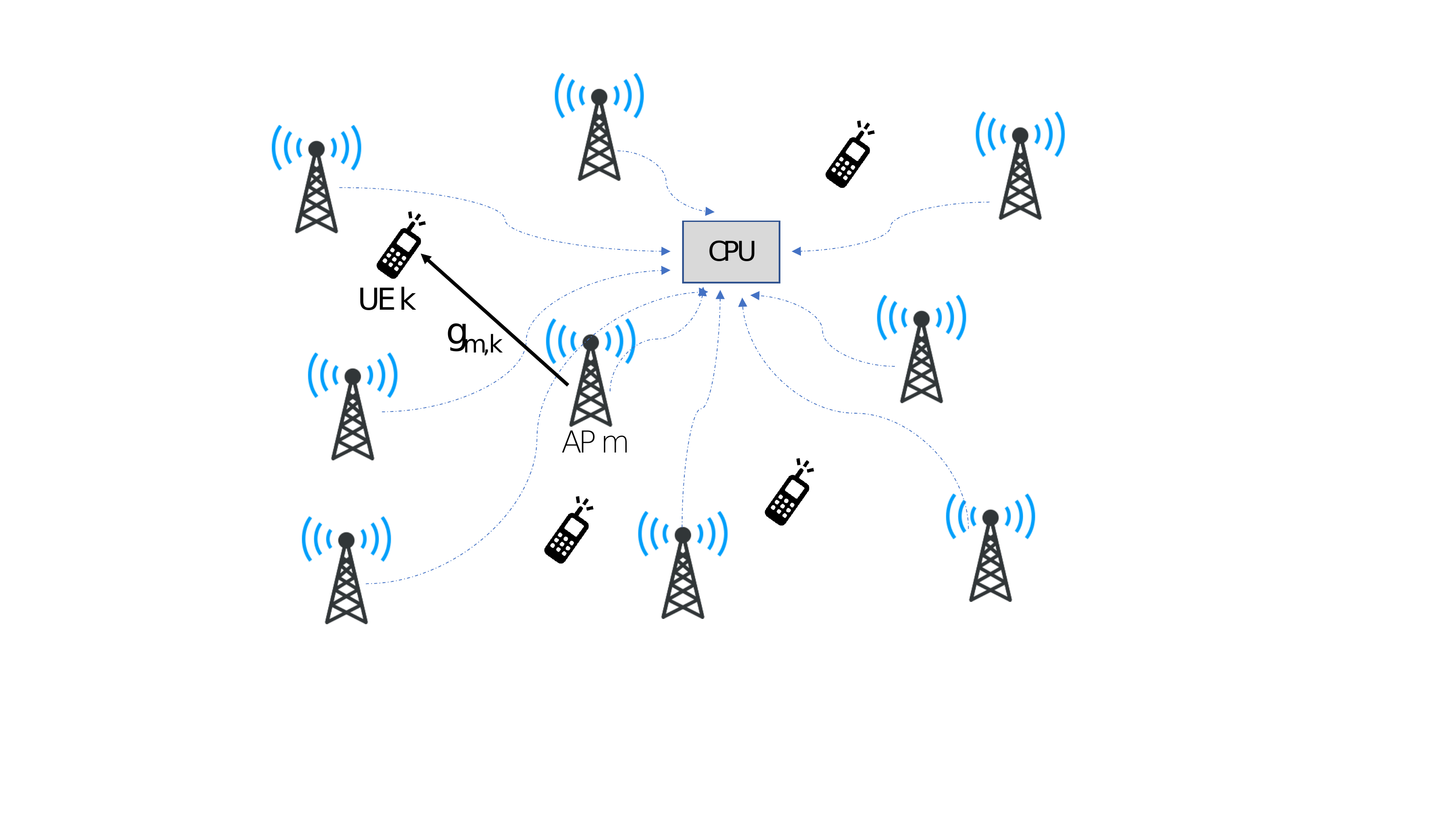} 
	\caption{In a cell-free Massive MIMO system, there are many distributed APs that are jointly serving the users. The encoding/decoding of signals can take place at a CPU. Each AP can be equipped with one or multiple antennas.}
\end{subfigure}
\caption{Comparison between cellular and cell-free Massive MIMO systems.}
\label{fig:system-model}
\end{figure}

CF Massive MIMO is essentially a single-cell Massive MIMO system with antennas distributed over a wide geographical area; see Fig.~\ref{fig:system-model}(b).
Hence, the joint channel from the APs to a user is strongly spatially correlated---some APs are closer to the user than others \cite{SIG-093}. The capacity of Massive MIMO systems with spatially correlated channels and imperfect CSI has been extensively analyzed in \cite{Huh2012a,Hoydis2013a,Yin2013a,SIG-093}. However, these prior works use channel models designed for co-located antenna arrays, which provable provide both channel hardening and  favorable propagation.
It is claimed in \cite{CF_mimo} that the outstanding aspect of CF Massive MIMO is that it can also utilize these phenomena, but this has not been fully demonstrated so far. Hence, it is not clear how to measure the achievable performance---is it reasonable to use the conventional Massive MIMO capacity lower bounds that rely on channel hardening or will they severely underestimate the CF performance? For example, \cite{interdonato2016much} derived a new downlink capacity lower bound for the case when the precoded channels are estimated using downlink pilots. That bound provides larger values than the common bound that estimates the precoded channels by relying on channel hardening, but it is unclear whether this indicates the need for downlink pilots and/or the lack of channel hardening in the CF Massive MIMO setup; one can alternatively estimate the effective channel blindly from the received signal.

While the basic signal processing techniques for precoding, combining, and channel estimation can be reused from the Massive MIMO literature, the distributed nature of CF Massive MIMO systems makes many ``simple'' resource allocation tasks highly nontrivial. Scheduling, power control, pilot allocation, system information broadcast, and random access are functionalities that are conventionally carried out on a per-cell basis, but these functionalities need to be implemented in a distributed fashion in CF Massive MIMO due to the lack of cells. These are important open issues that need to be treated in future work \cite{Interdonato2019a}, but we first need to understand to what extent we can rely on channel hardening and  favorable propagation to simplify these functionalities.

\subsection{Problem Statement and Methodology}

To enhance the understanding of the basic properties of CF Massive MIMO, this paper aims at answering the following open questions:
\begin{enumerate}
\item Can we observe channel hardening and favorable propagation in CF Massive MIMO with single-antenna APs?
\item Is it more beneficial to deploy more antennas on few APs or more APs with few antennas, in order to achieve a reasonable degree of channel hardening and favorable propagation?
\item What are the practical important factors that affect the conditions of these two properties?
\item Which capacity bounds from the conventional cellular Massive MIMO literature are appropriate to use in CF Massive MIMO?
\end{enumerate}
In order to answer these questions, we model the AP distribution by a homogeneous Poisson Point Process (PPP), where each AP is equipped with $N\geq 1$ of antennas. Unlike the conventional regular grid model for the base station deployment, the stochastic point process model considered in this work can capture the irregular and semi-random AP deployment in real networks \cite{lu2015stochastic, andrews_tractable}. Since a given user only observes one realization of the AP locations, we cannot treat these locations as random when evaluating the channel hardening and favorable propagation.
\textit{Hence, by first conditioning on a specific network realization with APs located at fixed locations and a reference user point at the origin}, we define the channel hardening and favorable propagation criteria as functions of the AP-user distances.
Then, we examine the spatially averaged percentage/probability of randomly located users that satisfy these criteria in a network with random AP locations.
The separation of the randomness caused by small-scale fading and by the spatial locations of APs is similar to the concept of \textit{meta distribution} proposed in \cite{meta_haenggi}, where the difference mainly lies in the definition of the studied performance metrics. 

Our analysis is carried out by considering different number of antennas per AP and different non-singular pathloss models: the single-slope model with different pathloss exponents \cite{ganti-adhoc, haenggi2009interference} and the multi-slope model \cite{CF_mimo}. Compared to the conference paper \cite{conference-version}, which focuses on the channel hardening aspect of CF Massive MIMO, in this paper, we provide a thorough investigation of both channel hardening and favorable propagation, based on which we also give insights into the selection of achievable rate expressions in CF Massive MIMO.

The remainder of this paper is organized as follows.
In Section~\ref{sec:system} we describe the CF Massive MIMO network model, including the AP distribution and the channel models. Next, 
Section~\ref{sec:harden} analyzes the channel hardening and Section~\ref{sec:favorable} analyzes the favorable propagation in CF Massive MIMO. Section~\ref{sec:capacity-bounds} considers different capacity lower bounds from cellular Massive MIMO and demonstrates which ones are useful in CF systems. Section~\ref{sec:conclusion} concludes the paper and summarizes the answers to the four main questions stated above.

\section{System Model}
\label{sec:system}
We consider a CF Massive MIMO system in a finite-sized network region $\mathcal{A}$. The APs are distributed on the two-dimensional Euclidean plane according to a homogeneous PPP $\Phi_A$ with intensity $\lambda_A$, measured in APs per m$^2$ \cite{haenggi2012stochastic}. Each AP is equipped with $N\geq 1$ antennas, which is a generalization of the $N=1$ CF Massive MIMO considered in prior works \cite{Ngo2015a,Nayebi2015a,CF_mimo,Nayebi2017a}. All the APs in the network are connected to a central processing unit (CPU) through backhaul, and the CPU codes and decodes the data signals; see Fig.~\ref{fig:system-model} for an illustration and \cite{Ngo2015a,Nayebi2015a,CF_mimo,Nayebi2017a,SIG-093} for the basic implementation aspects. Different to a small-cell network, all the APs are coordinated to serve all users simultaneously using the same frequency-time resources.
The number of users and their locations are generated by another independent point process. Denote by $L$ the number of APs in a specific realization of the PPP $\Phi_A$, we have that $L$ is a Poisson random variable (RV) with mean value
\begin{equation}
\mathbb{E}[L]=\lambda_A S(\mathcal{A}),
\end{equation}
where $S(\mathcal{A})$ denotes the area of the network region  $\mathcal{A}$.
Let the RV $M$ denote the total number of antennas existing in $\mathcal{A}$, then we have $M=LN$ and $\mathbb{E}[M]=N\lambda_A S(\mathcal{A})$.

Given the user distribution, we assume that there are $K$ users in a specific network realization, where the densities are selected to make $K\ll M$ in most realizations. Consider a typical user as a randomly chosen user inside the network region. When the AP density is much larger than the user density, the boundary effect caused by the finite-size network region is weak, i.e., users located at the network boundary are still likely to have nearby dominant APs that makes their received signal distribution similar to network-center users \cite{power_control_andrews}. Moreover, in the analysis, we will later let the network region grow infinitely large to alleviate any boundary effect. Thus, in the remainder of this paper, we consider having the typical user at the origin. The spatially averaged network statistics seen at this typical user can represent the average network performance seen by randomly located users in the network.

Denote by $\mathbf{g}_k$ the $M\times 1$ channel vector between all the antennas and the typical user (labeled as user $k$), the $m$-th element $g_{m, k}$ is modeled by
\begin{equation}
g_{m,k}=\sqrt{l(d_{m,k})} h_{m,k}, 
\label{eq:channel-vector}
\end{equation}
where $h_{m,k}$ represents the small-scale fading and $l(d_{m,k})$ represents the distance-dependent pathloss and it is a function of the distance $d_{m,k}$ between the $m$-th antenna and the user $k$. Since every $N$ antennas are co-located at the same AP, we have $d_{(i-1)\cdot N+1,k}=d_{(i-1)\cdot N+2,k}=\ldots=d_{i\cdot N,k}$, for $i=1,\ldots, L$.

As all previous works on cell-free Massive MIMO \cite{Ngo2015a,Nayebi2015a,CF_mimo,Nayebi2017a,Bashar2018a,interdonato2016much}, we assume independent Rayleigh fading from each antenna to the typical user, which means that $\{h_{m,k}\}$ are independently and identically distributed (i.i.d.) $\mathcal{CN}(0,1)$ RVs. The independence is natural for spatially separated antennas and the Rayleigh fading models a rich scattering environment. In the first part of this paper, we consider a non-singular pathloss model $l(r)=\min(1, r^{-\alpha})$, where $r$ is the antenna-user distance and $\alpha>1$ is the pathloss exponent.\footnote{Note that the unbounded pathloss model $l(r)= r^{-\alpha}$ is not appropriate when analyzing CF Massive MIMO with stochastic geometry, because the antennas can then be arbitrarily close to the user, which might result in unrealistically high power gain when using the unbounded pathloss model.} A three-slope pathloss model will also be studied in Section~\ref{sec:threeslope}.

Note that we do not include shadow fading in our analysis. The randomness caused by shadow fading coefficients can be seen as displacing the APs and varying the distances between the APs and the users. Whether to include shadow fading or not changes the exact values of the channel gain distribution, but the general trends of channel variation and orthogonality with respect to the antenna density will not be affected.\footnote{In the following sections, simulation results will be provided to support this claim.} Therefore, the inclusion of shadowing coefficients would not change the main conclusions of this paper.

\subsection{Main Advantage of CF Massive MIMO}
\label{sec:channel_spatial}
Similar to other distributed antenna systems, the main advantage of CF Massive MIMO is the macro-diversity; that is, the reduced distance between a 
user and its nearest APs. This can be demonstrated by analyzing the distribution of the squared norm of the channel vector,
\begin{equation} \label{eq:gamma-squared-norm}
\|\mathbf{g}_k\|^2=\sum\limits_{m=1}^{M} |h_{m,k}|^2 l(d_{m,k}),
\end{equation}
which we refer to as the \textit{channel gain}, since this is the gain of the effective scalar channel when using maximum-ratio (MR) precoding/combining. Note that $M$ depends on the realization of the PPP $\Phi_A$.

Each AP is equipped with $N$ antennas and therefore the sum of the squared magnitudes  $|h_{m,k}|^2$ of the small-scale fading coefficients of its $N$ co-located antennas is a RV following a $\text{Gamma}(N,1)$-distribution, which has mean $N$ and variance $N$. We define the distance vector $\mathbf{r}=[r_{1},\ldots, r_{L}]^T$, where each element $r_{i}$ denotes the distance from the $i$-th AP to the typical user at the origin. Thus, the squared norm in \eqref{eq:gamma-squared-norm} can be written as
\begin{equation} \label{eq:gamma-squared-norm-2}
\|\mathbf{g}_k\|^2=\sum\limits_{i\in\Phi_A} H_{i} l(r_{i}),
\vspace{-0.2cm}
\end{equation} 
where $H_{i}=\!\!\!\sum\limits_{m=(i-1)\cdot N+1}^{i\cdot N}\!\!|h_{m,k}|^2 \sim \text{Gamma}(N,1)$ and $r_i=d_{(i-1)\cdot N+1,k}=\ldots=d_{i\cdot N,k}$ for $i=1,\ldots,L$.

Note that there are two sources of randomness in \eqref{eq:gamma-squared-norm-2}: $\{H_{i}\}$ and $\Phi_A$. When studying the channel distribution for a randomly located user, it is natural to consider the distribution of $\|\mathbf{g}_k\|^2$ with respect to both sources of randomness.
From prior studies on the application of stochastic geometry in wireless networks, it is well known that the sum of the received power from randomly distributed nodes is described by a shot noise process. The mean and variance of $\|\mathbf{g}_k\|^2$ averaging over the spatial distribution of the antennas are known for the unbounded and bounded pathloss models \cite{haenggi2009interference}. 

\begin{lemma} \label{lemma1}
	For our considered pathloss model $l(r)=\min(1, r^{-\alpha})$, in a finite network region with radius $\rho$ centered around the typical user, we have
	\begin{align}
	\mathbb{E}\left[\|\mathbf{g}_k\|^2\right]= \left\{
	\begin{array}{rcl}
	&N\lambda_A\pi\left(1+\frac{2\left(1-\rho^{2-\alpha}\right) }{\alpha-2}\right) & \text{if}~~\alpha\neq 2\\
	&N\lambda_A\pi\left(1+2\ln(\rho) \right)& \text{if}~~\alpha=2\\
	\end{array} \right.\label{eq:mean_g}
	\end{align}
	\begin{equation}
	\mathrm{Var}\left[\|\mathbf{g}_k\|^2\right]=(N^2+N)\lambda_A\pi\left(1+\frac{1-\rho^{2-2\alpha}}{\alpha-1}\right). \label{eq:var_g}
	\end{equation}
\end{lemma}
\begin{IEEEproof}
	\textnormal{See Appendix \ref{appen1}.}
\end{IEEEproof}
The mean values in Lemma~\ref{lemma1} can be applied to network regions of arbitrary size. For $\alpha>2$, $\mathbb{E}\left[\|\mathbf{g}_k\|^2\right]$ approaches the limit $N\lambda_A\pi\frac{\alpha }{\alpha-2}$ as $\rho\rightarrow\infty$, which demonstrates the fact that only a limited number of APs closest to the user have a non-negligible impact on the channel gain. In contrast, when $1<\alpha<2$, both the mean and variance of $\|\mathbf{g}_k\|^2$ increases unboundedly when the network region grows.

It is particularly interesting to study the case when the antenna density $\antdensity=N\lambda_A$ is fixed. We then observe that the average channel gain in \eqref{eq:mean_g} is the same, irrespective of whether there is a high density of single-antenna APs or a smaller density of multi-antenna APs. The variance is, however, proportional to $(N^2+N)\lambda_A = (N+1)\antdensity$ and thus grows with $N$.
\begin{figure}[h]
	\centering
	\includegraphics[width=0.5\textwidth]{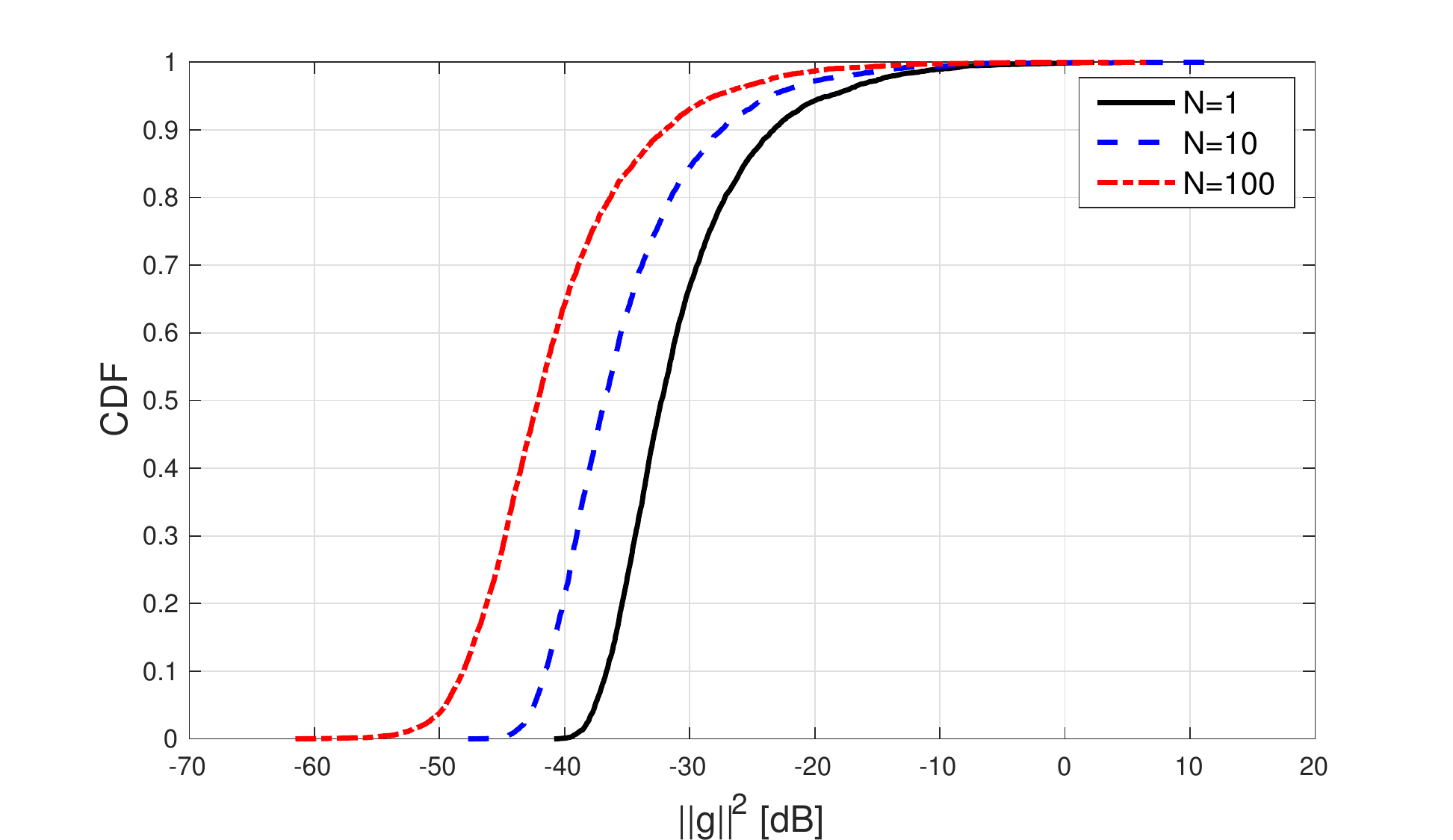}
	\caption{The CDF of the squared norm of the channel vector $\|\mathbf{g}_k\|^2$ to user $k$ with respect to small-scale fading and PPP realizations. The number of antennas per AP is $N=\{1, 10, 100\}$. The antenna density is fixed $\antdensity=N\lambda_A=0.001$/m$^2$ ($10^3$/km$^2$). }
	\label{fig:cdf_channel_vector}
\end{figure}

Fig.~\ref{fig:cdf_channel_vector} demonstrates a key benefit of CF Massive MIMO. More precisely, the figure shows the cumulative distribution function (CDF) of $\|\mathbf{g}_k\|^2$, with respective to random spatial locations and small-scale fading realizations. We consider $\antdensity=10^3$/km$^2$ and different numbers of antennas per AP: $N \in \{1,\,10,\,100\}$. Note that the horizontal axis is shown in decibel, thus a few users are very close to an AP and have large values of $\|\mathbf{g}_k\|^2$ while the majority have substantially smaller values.
The long-tailed exponential distribution of the small-scale fading $|h_{m,k}|^2$ has a strong impact on the CDF, but also the AP density makes a difference. The larger $N$ is, the longer tail the distribution has; at the 95\%-likely point, $N=1$ achieves a 12\,dB higher value than $N=100$.
The reason behind the increasing tail with $N$ is that $\text{Var}[\|\mathbf{g}_k\|^2]$ is proportional to $(N+1)$, as described above. A practical interpretation is that having higher AP density reduces the average distance between the APs and this macro-diversity reduces the risk that a randomly located user has large distances to all of its closest APs. This is a key motivation for deploying CF Massive MIMO systems with $N=1$ and high AP density; for example, the system will provide a more uniform coverage to users at random locations than a conventional cellular Massive MIMO deployment with $N=100$ and a low AP density. This property should be kept in mind when we later show the downsides with having a high AP density.

From \cite{haenggi2009interference} and \cite{interference}, it is known that a Gamma distribution provides a good approximation of the interference distribution in a Poisson random field with non-singular pathloss. Here, the expression of $\|\mathbf{g}_k\|^2$ coincides with the definition of interference power in \cite{haenggi2009interference} and \cite{interference}. Thus, the Gamma distribution can be used to approximate the distribution of $\|\mathbf{g}_k\|^2$. The details are omitted since it is outside the scope of this work.

\subsection{Conditional Channel Distribution at Fixed Location}
\label{sec:channel_conditional}
The previous analysis characterized the channel gain distribution that a user will observe when moving around in a large network.
Once the APs are deployed, for a user at a fixed location (e.g., located in a room), the small-scale fading varies over time but the large-scale fading from the APs to the user remains the same. Conditioning on a specific network realization of $\Phi_A$, assuming that there are $L$ APs in the network, the distances between the APs and the typical user are basically fixed. The conditional distribution of the channel statistics with respect to the small-scale fading distribution is essential for performance evaluation (e.g., computing the ergodic capacity) of CF Massive MIMO networks with a fixed topology and users at fixed but randomly different locations. 

With single-antenna APs, i.e., $N=1$, the total number of antennas $M$ is equal to the number of APs $L$. As a result of the exponentially distributed small-scale fading coefficient $|h_{m,k}|^2$, the conditional distribution of the channel gain  $\|\mathbf{g}_k\|^2$ in \eqref{eq:gamma-squared-norm} follows a Hypoexponential distribution, denoted by $\mathrm{Hypo}(l(r_{1})^{-1}, \ldots, l(r_{L})^{-1})$, which is usually a long-tailed distribution when the coefficients $l(r_{i})$ are distinct \cite{hypoexpo}.

With $N>1$ antennas per AP, the channel gain $\|\mathbf{g}_k\|^2$ is given by \eqref{eq:gamma-squared-norm-2}
where $H_{i} \sim \text{Gamma}(N,1)$. As a result, the conditional distribution of the channel gain from the $i$-th AP is $H_{i} l(r_{i}) \sim \text{Gamma}(N, l(r_{i}))$ for  $i=1,\ldots,L$. Due to the sum of independent Gamma RVs with different scale parameters, the mean and variance of $\|\mathbf{g}_k\|^2$ conditioning on the distance vector $\mathbf{r}$ are
\begin{align}
\mathbb{E} \left[\|\mathbf{g}_k\|^2 \Bigm| \mathbf{r} \right]&=  N\sum\limits_{i=1}^{L} l(r_{i})  \label{mean_condi}\\
\text{Var}\left[\|\mathbf{g}_k\|^2 \Bigm| \mathbf{r} \right]&=N\sum\limits_{i=1}^{L} l^2(r_{i}). \label{var_condi}
\end{align}
The exact conditional probability density function (PDF) of $\|\mathbf{g}_k\|^2$ can be computed using the approach in \cite{Amari1997a} and the exact expression is available in \cite[Eq.~(6)]{Bjornson2009c}. By looking at \eqref{mean_condi} and \eqref{var_condi}, it is unclear how the channel gain behaves; for example, if it is the mean value or the channel variations that grow faster.

In cellular Massive MIMO with $M$ co-located antennas, conditioning on a specific location of the user, denote by $\beta=\mathbb{E}[ \|\mathbf{g}_k\|^2 | \mathbf{r}]/M$ the pathloss from the $M$ co-located antennas to the user. The squared norm of the channel gain $\|\mathbf{g}_k\|^2$ then follows a $\text{Gamma}(M,\beta)$-distribution, with mean value $\beta M$ and standard deviation $\beta \sqrt{M}$. When $M$ increases, the distribution approaches a normal distribution and is (relatively speaking) concentrated around the mean since it grows faster than the standard deviation. The different channel gain distributions in CF and cellular Massive MIMO highlight the fundamental difference between the channel statistics of these two types of networks. In the remainder of this paper, we will proceed to investigate if in a CF Massive MIMO network, we could observe the classical Massive MIMO phenomena, namely channel hardening and favorable propagation.

\section{Measure of Channel Hardening}
\label{sec:harden}
In cellular Massive MIMO, when the number of antennas grows, the channel between the AP and the user behaves as almost deterministic. This property of is called \textit{channel hardening}. 
Conditioning on a specific network realization with distance vector $\mathbf{r} =[r_{1},\ldots,r_{L}]^T$, channel hardening appears for the typical user in CF Massive MIMO when the following condition holds:
\vspace{-0.1cm}
\begin{equation}
\frac{\|\mathbf{g}_k\|^2}{\mathbb{E}\left[\|\mathbf{g}_k\|^2 | \mathbf{r}  \right]}\rightarrow 1~~\text{as}~~ M\rightarrow \infty,
\vspace{-0.2cm}
\label{eq:convergence_channel_gain}
\end{equation}
where $\|\mathbf{g}_k\|^2=\sum\limits_{i=1}^{L}  H_{i} l(r_{i})$ is the channel gain from the $L$ APs to the typical user. 
One way to prove channel hardening (with convergence\footnote{Note that convergence in mean square implies convergence in probability.} in \eqref{eq:convergence_channel_gain} in mean square sense) is to show that the channel gain variation
\vspace{-0.1cm}
\begin{equation}
\text{Var} \left[ \frac{\|\mathbf{g}_k\|^2}{\mathbb{E}\left[\|\mathbf{g}_k\|^2| \mathbf{r}  \right]} \bigg| \mathbf{r} \right] = \frac{\text{Var}\left[\|\mathbf{g}_k\|^2 | \mathbf{r} \right]}{\left(\text{E}\left[\|\mathbf{g}_k\|^2 | \mathbf{r} \right]\right)^2} \to 0 ~~\text{as}~~ M\rightarrow \infty.
\end{equation}

For a large wireless network, studying the channel statistics at a specific location is of limited interest and the results cannot be generalized to users at other arbitrary locations.
To quantify the channel gain variation for users at arbitrary locations, we define the following channel hardening measure:
\begin{equation}
p_{\theta}=\mathbb{P}\left[\frac{\text{Var}\left[\|\mathbf{g}_k\|^2\Bigm| \mathbf{r}\right]}{\left(\text{E}\left[\|\mathbf{g}_k\|^2\Bigm| \mathbf{r}\right]\right)^2}\leq \theta\right].
\label{eq:ptheta}
\end{equation}
This is the CDF of $\frac{\text{Var}\left[\|\mathbf{g}_k\|^2 | \mathbf{r}\right]}{\left(\text{E}\left[\|\mathbf{g}_k\|^2 | \mathbf{r}\right]\right)^2}$ given a certain threshold $\theta$. 
Here, the probability is obtained over different network realizations that generate different distance vector $\mathbf{r}$. As mentioned in Section~\ref{sec:system}, the spatially averaged probability $p_{\theta}$ provides the percentage of randomly located users that experience $\frac{\text{Var}\left[\|\mathbf{g}_k\|^2 | \mathbf{r}\right]}{\left(\text{E}\left[\|\mathbf{g}_k\|^2 | \mathbf{r}\right]\right)^2}$ smaller or equal to $\theta$. Notice that $p_{\theta}=1$ implies that all users have channel gain variations that are smaller or equal to $\theta$. The ideal case is
$p_{0}=1$ where the variance is zero for all users. When the threshold $\theta$ is small enough, the larger $p_{\theta}$ is, with higher possibility we observe channel hardening for users at arbitrary locations.

\subsection{Necessary Conditions for Channel Hardening}
\label{sec:harden_multiple}
With $N\geq 1$ antennas per AP, from \eqref{mean_condi} and \eqref{var_condi}, the channel hardening measure in \eqref{eq:ptheta} can be written as
\begin{equation}
p_{\theta}= \mathbb{P}\left[\frac{N\sum_{i=1}^{L}l^2(r_i)}{\left(N\sum_{i=1}^{L}l(r_i)\right)^2} \leq \theta\right] =\mathbb{P}\left[\frac{\sum_{i=1}^{L}l^2(r_i)}{N\left(\sum_{i=1}^{L}l(r_i)\right)^2} \leq \theta\right]. 
\label{eq:ptheta_N}
\end{equation}
Since $N$ appears in the denominator, for a given $\theta$, $p_\theta$ always increases with $N$.  This implies that regardless of the AP density, having more antennas per AP always helps the channel to harden. In the following, we fix the number of antennas $N$ per AP and study the impact that the AP density $\lambda_A$ has on the channel hardening criterion.

For a given network realization with $L$ APs, by defining $Y_{1}=\sum_{i=1}^{L}l(r_i)$, $Y_{2}=\sum_{i=1}^{L}l^2(r_i)$, and 
\begin{equation}
X_{\text{ch}}=\frac{Y_{2}}{N Y_{1}^2}=\frac{\sum_{i=1}^{L}l^2(r_i)}{N\left(\sum_{i=1}^{L}l(r_i)\right)^2},
\end{equation}
we can write the channel hardening measure as
\vspace{-0.2cm}
\begin{equation}
p_{\theta}=\mathbb{P}\left[X_{\text{ch}} \leq \theta\right].
\vspace{-0.15cm}
\end{equation}

The exact of distribution of $X_{\text{ch}}$ is hopeless to analyze, even with the joint PDF of $r_i$, $i=1,\ldots,L$, because $Y_{1}$ and $Y_{2}$ are strongly correlated. One objective of this work is to provide intuitive insights into the relation between channel hardening and the AP density, without relying on extensive numerical simulations. Therefore, instead of studying the original channel hardening measure in \eqref{eq:ptheta_N}, we will obtain a strongly related measure that asymptotically follows the same trends as $p_{\theta}$, but is more analytical tractable.

Specifically, if $p_\theta$ should approach $1$ when the AP density $\lambda_A$ increases, we need $X_{\text{ch}}=\frac{Y_{2}}{N Y_{1}^2}\rightarrow 0$ when $\lambda_A\rightarrow\infty$.  
Though the distributions of $Y_2$ and $Y_1^2$ are not trivial to obtain, their mean and variance can be obtained by Campbell's theorem as in Section~\ref{sec:channel_spatial}. 

\begin{lemma}
	\label{lemma2}
	For the non-singular pathloss model $l(r)=\min(1, r^{-\alpha})$, in a network region with radius $\rho$, we have
	\begin{align}
	\mathbb{E}\left[Y_1\right]= \left\{
	\begin{array}{rcl}
	&\!\!\!\!\lambda_A\pi\left(1+\frac{2\left(1-\rho^{2-\alpha}\right) }{\alpha-2}\right) & \text{if}~~\alpha\neq 2\\
	&\!\!\!\!\lambda_A\pi(1+2\ln (\rho) )& \text{if}~~\alpha=2
	\end{array} \right.
	\label{eq:mean_Y1}
	\end{align}
	\begin{equation}
	\textnormal{Var}\left[Y_1\right]=\lambda_A\pi\left(1+\frac{1-\rho^{2-2\alpha}}{\alpha-1}\right).
	\end{equation}
\end{lemma}
\begin{IEEEproof}
	\textnormal{See Appendix \ref{appen2}.}
\end{IEEEproof}

Then, using $\mathbb{E}[Y_{1}^2]=\text{Var}[Y_1]+\left(\mathbb{E}[Y_1]\right)^2$, we obtain
\begin{align}
\mathbb{E}\left[Y_{1}^2\right]= \left\{
\begin{array}{rcl}
&\!\!\!\!\!\lambda_A\pi\left(\frac{\alpha-\rho^{2-2\alpha} }{\alpha-1}+\lambda_A\pi\left(\frac{\alpha-2\rho^{2-\alpha}}{\alpha-2}\right)^2\right)& \text{if}~~\alpha\neq 2\\
&\!\!\!\!\!\lambda_A\pi\left(\frac{\alpha-\rho^{2-2\alpha} }{\alpha-1}+\lambda_A\pi\left(1+2\ln (\rho) \right)^2\right)& \text{if}~~\alpha=2.\\
\end{array} \right.
\end{align}
For $Y_{2}=\sum_{i=1}^{L}l^2(r_i)$, using again Campbell's theorem, we have
\begin{equation}
\mathbb{E}[Y_{2}]= \lambda_A\pi\frac{\alpha-\rho^{2-2\alpha} }{\alpha-1}=\text{Var}[Y_{1}].
\vspace{-0.2cm}
\end{equation}
From the above results, we make the following observations:
\begin{itemize}
	\item $Y_2$ scales proportionally to $\lambda_A$;
	\item The higher order element of $Y_1^2$ scales proportionally to $\lambda_A^2$;
	\item When the pathloss is bounded, both $Y_2$ and $Y_1^2$ have finite mean, which increase with $\lambda_A$.
\end{itemize}
Given that both $Y_2$ and $Y_1^2$ increase with $\lambda_A$, if we need $\frac{Y_{2}}{NY_{1}^2}\rightarrow 0$ when $\lambda_A$ increases, then $Y_1^2$ should grow faster than $Y_2$. Hence, we need $\frac{\mathbb{E}[Y_{2}]}{N\mathbb{E}[Y_{1}^2]}\rightarrow 0$ when $\lambda_A$ increases. In other words, if $\frac{\mathbb{E}[Y_{2}]}{N\mathbb{E}[Y_{1}^2]}$ does not converge to zero when $\lambda_A$ increases, adding more APs will not help the channel to harden. We continue to investigate this necessary condition for channel hardening below.

When the network region grows infinity large, i.e., $\rho\rightarrow \infty$, depending on the pathloss exponent, we have the following cases:
\subsubsection{$\alpha>2$}
As the network radius $\rho\rightarrow\infty$, we have $\rho^{2-2\alpha}\rightarrow 0$ and $\rho^{2-\alpha}\rightarrow 0$, which implies
\vspace{-0.2cm}
\begin{eqnarray}
\mathbb{E}\left[Y_{1}^2\right]&\rightarrow&\lambda_A\pi\left(\frac{\alpha }{\alpha-1}+\lambda_A\pi\left(\frac{\alpha}{\alpha-2}\right)^2\right),\\
\vspace{-0.2cm}
\mathbb{E}[Y_{2}]&\rightarrow& \lambda_A\pi\frac{\alpha}{\alpha-1},\\
\vspace{-0.2cm}
\frac{\mathbb{E}[Y_{2}]}{N \mathbb{E}\left[Y_{1}^2\right]}&\to&\frac{1/N}{1 + \lambda_A\pi\frac{\alpha  (\alpha-1)}{\left(\alpha-2\right)^2}} .\label{channel_N}
\end{eqnarray}
With small $N$, in order to have $\frac{\mathbb{E}[Y_{2}]}{N\mathbb{E}\left[Y_{1}^2\right]}$ approaching $0$, the AP density should satisfy $\lambda_A\frac{\alpha \pi (\alpha-1)}{\left(\alpha-2\right)^2}\gg 1$. Since the AP density is measured in APs per m$^2$, the condition for channel hardening is only satisfied if  $\lambda_A \sim 1 $ AP/m$^2$, which is a rather unrealistic condition in practice.

\subsubsection{$\alpha=2$}
This case behaves as in a free-space propagation environment. As $\rho\rightarrow \infty$, we have $\ln (\rho) \rightarrow \infty$ and $\rho^{2-2\alpha}\rightarrow 0$, which implies
\begin{eqnarray}
\!\!\!\!\mathbb{E}[Y_2]&\!\!\!\!\rightarrow\!\!\!\!&\lambda_A\pi\frac{\alpha}{\alpha-1}, \\
\!\!\!\!\frac{\mathbb{E}[Y_{2}]}{N \mathbb{E}\left[Y_{1}^2\right]}&\!\!\!\!\asymp\!\!\!\!&\frac{1/N}{1+\lambda_A\pi(1+2\ln (\rho) )^2\frac{\alpha-1}{\alpha }}\rightarrow 0,\label{alpha2_conditions}
\vspace{-0.15cm}
\end{eqnarray}
where the operator $\asymp$ means that the difference between the expressions vanishes asymptotically.
From \eqref{alpha2_conditions}, we observe that channel hardening is achieved as the network radius increases.

\subsubsection{$1<\alpha<2$}
One example of this case is the indoor near field propagation. With $\rho\rightarrow \infty$, we have $\rho^{2-\alpha}\rightarrow\infty$ and $\rho^{2-2\alpha}\rightarrow 0$, which implies
\begin{align}
\frac{\mathbb{E}[Y_{2}]}{N\mathbb{E}\left[Y_{1}^2\right]}\asymp \frac{1/N}{1+\lambda_A\pi\frac{4\rho^{4-2\alpha} }{(2-\alpha)^2}\frac{\alpha-1}{\alpha }}\rightarrow 0.\label{alpha12_conditions}
\vspace{-0.15cm}
\end{align}

From the above equations, we see that $\frac{\mathbb{E}[Y_2]}{N \mathbb{E}\left[Y_1^2\right]}$ decreases rapidly with $\lambda_A$ and $\rho$ when $\alpha\leq2$. When the network region grows infinitely large, $\frac{\mathbb{E}[Y_2]}{N \mathbb{E}\left[Y_1^2\right]}$ will eventually approach $0$. This suggests that with smaller pathloss exponents, e.g., free-space propagation and indoor near-field propagation, it is more likely to observe channel hardening in CF Massive MIMO. With the two-ray ground-reflection pathloss model and $\alpha=4$ \cite{goldsmith2005wireless}, the convergence to channel hardening only happens with impractically high antenna density.

\begin{figure}[h!]
	\centering
	\includegraphics[width=0.5\textwidth]{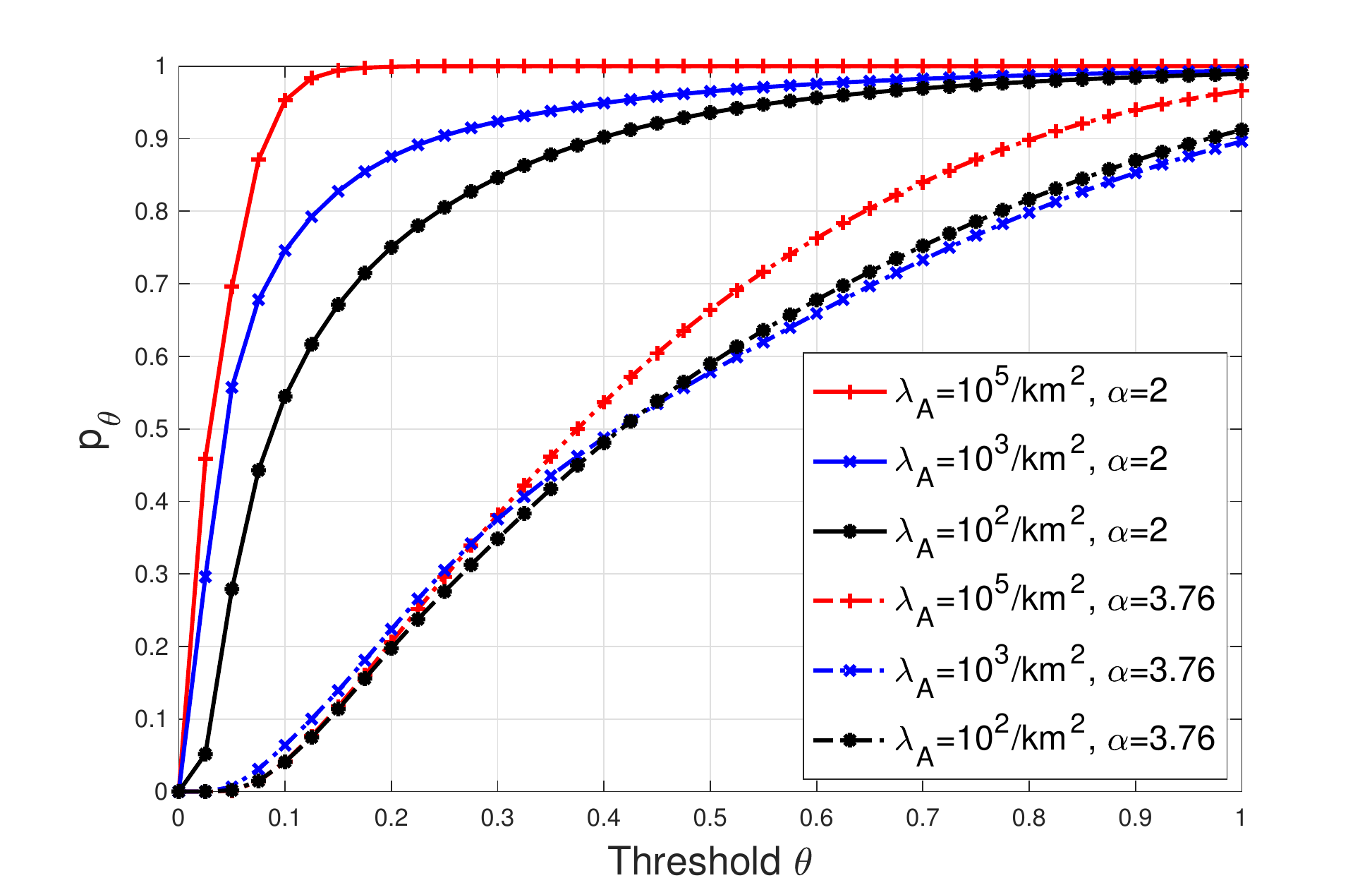}
	\caption{The CDF of $X_{\text{ch}}$, with pathloss exponent $\alpha\in\{3.76, 2\}$. The network radius is $\rho=0.5$ km and $N=1$. The AP density is $\lambda_A \in \{10^2, 10^3, 10^5\}/$km$^2$, which is equivalent to $\{10^{-4}, 10^{-3}, 0.1\}/$m$^2$. }
	\label{fig:test_channel}
\end{figure}

To validate our analytical predictions from \eqref{channel_N}, \eqref{alpha2_conditions} and \eqref{alpha12_conditions}, we present in Fig.~\ref{fig:test_channel} the simulated $p_\theta$  (i.e., the CDF of $X_{\text{ch}}$) for different AP densities, obtained with pathloss exponents $\alpha \in\{3.76, \, 2\}$. Note that in this figure we only consider $N=1$.
We have chosen large values of $\lambda_A$ in order to see the behavior of $p_\theta$ when $\lambda_A\rightarrow \infty$.
Fig.~\ref{fig:test_channel} shows that with $\alpha=3.76$, for a given threshold $\theta$, the channel hardening measure $p_\theta$ does not change much with the AP intensity, unless we reach $\lambda_A=10^{5}$/km$^2$ ($0.1$/m$^2$). However, having $\lambda_A>10^{3}$/km$^2$ is probably practically unreasonable. With $\alpha=2$, the convergence of the channel hardening measure $p_\theta$ to one becomes more obvious when the AP density grows, which indicates that the probability to observe channel hardening at random locations is fairly large. Since the simulation results match well our analytical predictions, it also validates that using $\frac{\mathbb{E}[Y_{2}]}{N\mathbb{E}[Y_{1}^2]}\rightarrow 0$ as the equivalent channel hardening condition is a reasonable choice.

\begin{figure}[h!]
	\centering
	\includegraphics[width=0.5\textwidth]{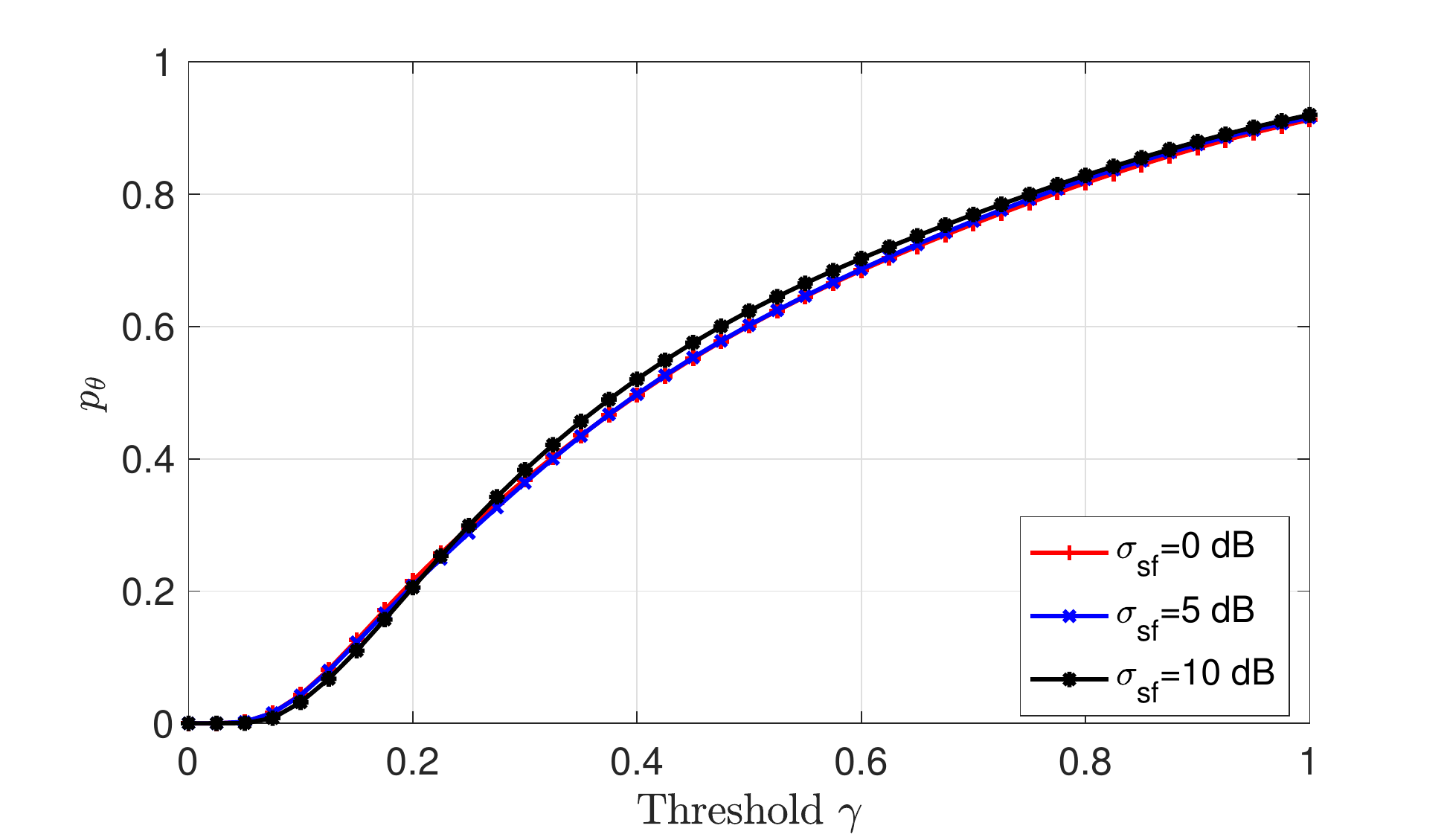}
	\caption{ The CDF of $X'_{\text{ch}}$ with log-normal shadowing fading. The standard deviation is $\sigma_{\text{sh}}=\{0, 5, 10\}$ dB. The AP density is $\lambda_A=10^{-4}$/m$^2$. The pathloss exponent is $\alpha=3.76$. The network radius is $\rho=0.5$ km and $N=1$.  }
	\label{fig:test_shadow}
	\vspace{-0.4cm}
\end{figure}
To validate our claim that the inclusion of shadow fading does not change our observations, in Fig.~\ref{fig:test_shadow}, we show the CDF of the channel hardening metric $X'_{\text{ch}}$ when including log-normal shadow fading. Here, we consider a similar shadowing model as in\cite{CF_mimo}, i.e., the large-scale fading coefficient from the $i$-th AP is denoted by $\beta_{i}=l(r_{i})\cdot 10^{\frac{\sigma_{\text{sh}} z_{i}}{10}}$, with $\sigma_{\text{sf}}$ being the standard deviation of the log-normal shadow fading and $z_{i}\sim\mathcal{N}(0,1)$ when $r_{i}>50$\,m. When $r_{i}\leq 50$\,m, there is no shadowing. Then the channel hardening metric is defined as $X'_{\text{ch}}=\frac{\sum_{i=1}^{L}\beta_i^2}{N\left(\sum_{i=1}^{L}\beta_i\right)^2}$.
In Fig.~\ref{fig:test_shadow}, the curves for $\sigma_{\text{sh}}=\{0, 5, 10\}$ dB almost overlap. Therefore, our conclusions on channel hardening will not be affected by the inclusion of shadowing in the channel model.

In summary, we have proved the following result.

\begin{theorem}
	Increasing the number of antennas per AP  in CF Massive MIMO always helps the channel to harden.
	With one antenna per AP, increasing the AP density does not lead to channel hardening when using typical pathloss exponents and AP densities. In a propagation environment with a very small pathloss exponent, $\alpha \leq 2$, the channel hardening criterion has higher chance to be satisfied as the AP density increases.
\end{theorem}

\subsection{More Antennas on Few APs or More APs with Few Antennas?}
Suppose the antenna density $\mu=N\lambda_A$ is fixed and we shall determine how to deploy these antennas. Whether a larger $N$ with smaller AP density $\lambda_A$ or vice versa gives a higher level of channel hardening can be inferred from \eqref{channel_N} for $\alpha>2$. 
We can rewrite \eqref{channel_N} as
\begin{equation}
\frac{\mathbb{E}[Y_{2}]}{N \mathbb{E}\left[Y_{1}^2\right]}=\frac{1}{N+N\lambda_A\pi\frac{\alpha  (\alpha-1)}{\left(\alpha-2\right)^2}}=\frac{1}{N+\mu \pi\frac{\alpha  (\alpha-1)}{\left(\alpha-2\right)^2}}.\label{eq:diff_N}
\end{equation} 
Since the denominator contains $N$ plus a constant term for fixed $\mu$, we will clearly obtain more channel hardening by having more antennas on fewer APs if the total amount of antennas is fixed.\footnote{This result was obtained with uncorrelated fading between the user and the antennas on an AP. If there instead is spatially correlated fading, due to insufficient scattering around the AP, this will slightly reduce the hardening, but more antennas will still be beneficial.} The same conclusions can be drawn from \eqref{alpha2_conditions} and \eqref{alpha12_conditions} for $\alpha\leq2$. Note that the stronger channel hardening comes at the price of less macro diversity. 

\begin{figure}[ht!]
	\vspace{-0.2cm}
	\centering
	\includegraphics[width=0.5\textwidth]{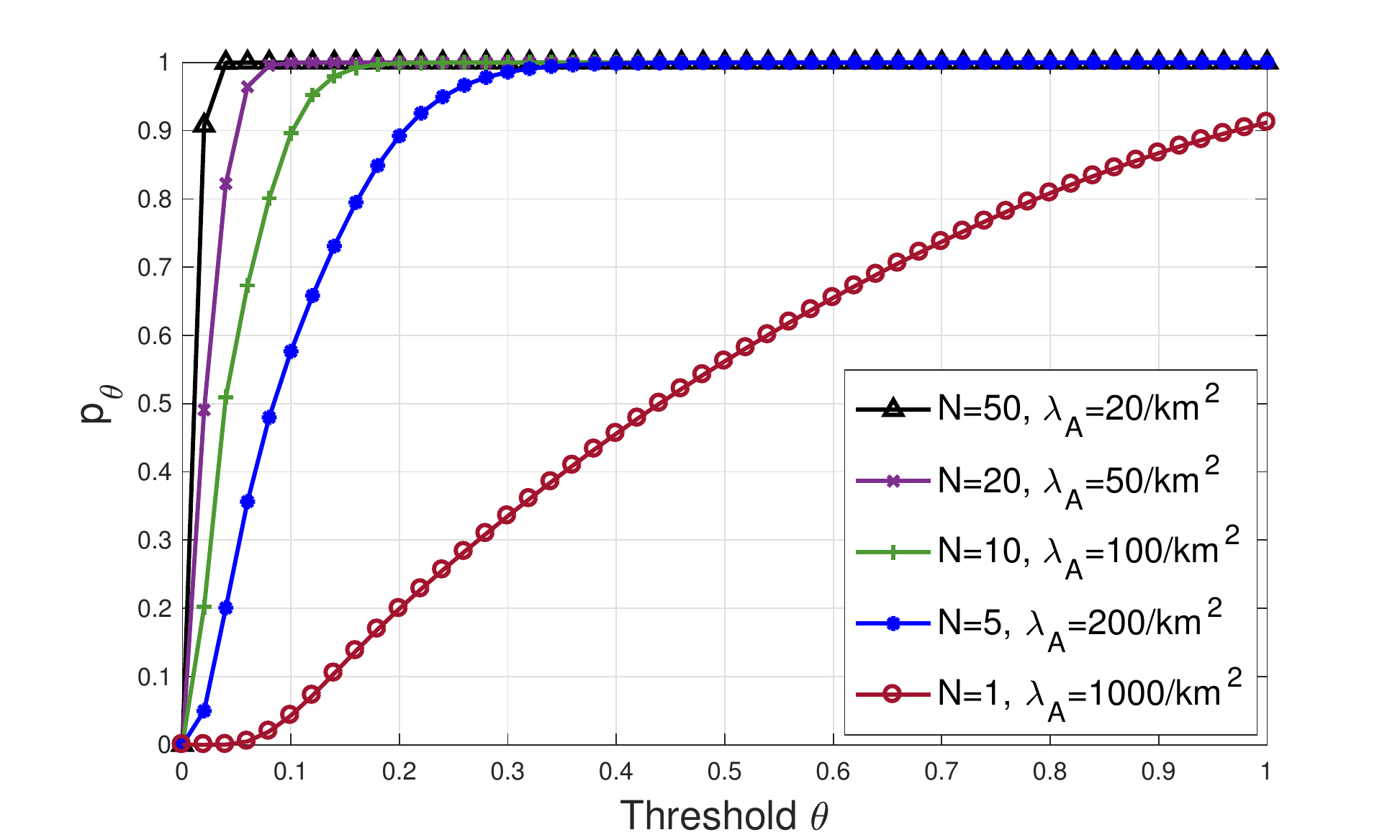}
	\vspace{-0.2cm}
	\caption{The CDF of $X_{\text{ch}}$ with pathloss exponent $\alpha=3.76$ and network radius $\rho=0.5$ km. The antenna density is $\mu = N\lambda_A=1000$/km$^2$ ($10^{-3}$/m$^2$).}
	\label{fig:ptheta_diff_N_fixed_M}
	\vspace{-0.2cm}
\end{figure}

To validate our analytical predictions, in Fig.~\ref{fig:ptheta_diff_N_fixed_M}, we present $p_\theta$ (i.e., the CDF of $X_{\text{ch}}$) for different $\lambda_A$ and $N$ while keeping the overall antenna density fixed at $\mu = N\lambda_A=10^3$/km$^2$ ($10^{-3}$/m$^2$). This figure confirms our prediction from \eqref{eq:diff_N} that having multiple antennas per AP will substantially help the channel to harden, and the level of channel hardening clearly increases with $N$. The curve $N=50$ can be interpreted as a cellular Massive MIMO system, due to the massive number of antennas per AP. 
The largest improvements occur when going from $N=1$ to $N=5$ (or to $N=10$), thus we can achieve reasonable strong channel hardening within the scope of CF Massive MIMO if each AP is equipped with an array of 5-10 antennas. With a smaller pathloss exponent, the required number of antennas per AP to achieve reasonably strong channel hardening is also smaller.

\subsection{Multi-Slope Pathloss Model} 
\label{sec:threeslope}
In this section, we extend our analysis to the scenario with a multi-slope pathloss model, which models the fact that the pathloss exponent generally increases with the propagation distance. 
Similar to \cite{CF_mimo}, we consider the three-slope pathloss model 
\begin{eqnarray}
l(r)= \left\{
\begin{array}{rcl}
Cr^{-3.5} & \text{if} & r> d_1\\
Cr^{-2} d_1^{-1.5}& \text{if} & d_0\leq r\leq d_1\\
Cd_0^{-2} d_1^{-1.5}& \text{if} & r< d_0,\\
\end{array} \right.
\label{eq:three-slope}
\end{eqnarray}
where $d_0$ and $d_1$ are fixed distances at which the slope starts to change, $C$ is a constant that depends on the carrier frequency and antenna height. Since the constant factor $C$ does not affect the channel hardening measure, for simplicity, we consider $C=1$ in the remainder of this section.

As previously in this paper, we consider $\frac{\mathbb{E}[Y_{2}]}{N\mathbb{E}\left[Y_{1}^2\right]}\rightarrow 0$ as the necessary condition of channel hardening. When $\rho\rightarrow\infty$, 
\begin{align}
&\mathbb{E}[Y_1] =\lambda_A 2\pi\int_{\mathbb{R}} l(r)  r\text{d}r 
=2\lambda_A\pi d_1^{-1.5}  \left(\ln (d_1)-\ln (d_0)+\frac{7}{6}\right), \\
&\text{Var}[Y_1] 
=\lambda_A 2\pi\int_{\mathbb{R}} l^2(r)   r\text{d}r 
=2\lambda_A\pi d_1^{-3}  \left(d_0^{-2}-\frac{3}{10}d_1^{-2}\right),
\end{align}
\begin{equation}
\begin{split}
\mathbb{E}\left[Y_{1}^2\right]=&\text{Var}[Y_1]+\left(\mathbb{E}[Y_1]\right)^2\\=&2\lambda_A\pi d_1^{-3}\!\!\left[2\pi\lambda_A\left(\ln( d_1)-\ln (d_0)+\frac{7}{6}\right)^2\!\!+\!d_0^{-2}\!-\!\frac{3}{10}d_1^{-2}\right]. 
\end{split}
\end{equation}

Since $\mathbb{E}[Y_2] =\text{Var}[Y_1]$, we have
\begin{equation}
\begin{split}
\frac{\mathbb{E}[Y_{2}]}{N \mathbb{E}\left[Y_{1}^2\right]}=&\frac{1/N}{1+2\lambda_A\pi \frac{\left(\ln (d_1)-\ln (d_0)+\frac{7}{6}\right)^2}{d_0^{-2}-\frac{3}{10}d_1^{-2}}} \\\approx&\frac{1/N}{1+2\lambda_A\pi d_0^2\left(\ln (d_1) -\ln (d_0)+\frac{7}{6}\right)^2},
\end{split}
\end{equation}
where the approximation holds when $d_1\gg 1$\,m.
Based on this, using larger $d_0$ and $d_1$ will make $\frac{\mathbb{E}[Y_{2}]}{N \mathbb{E}\left[Y_{1}^2\right]}$ approach $0$ with higher speed when $\lambda_A$ increases. 
This is intuitive, given the previous observation that a smaller pathloss exponent improves the hardening, because as $d_0$ and $d_1$ increase, the number of APs with small pathloss exponents increases. Adding more antennas to the APs will also improve the channel hardening, both for a fixed $\lambda_A$ and when the total antenna density $\mu = N \lambda_A$ is fixed.

\begin{figure}[ht!]
	\centering
	\includegraphics[width=0.5\textwidth]{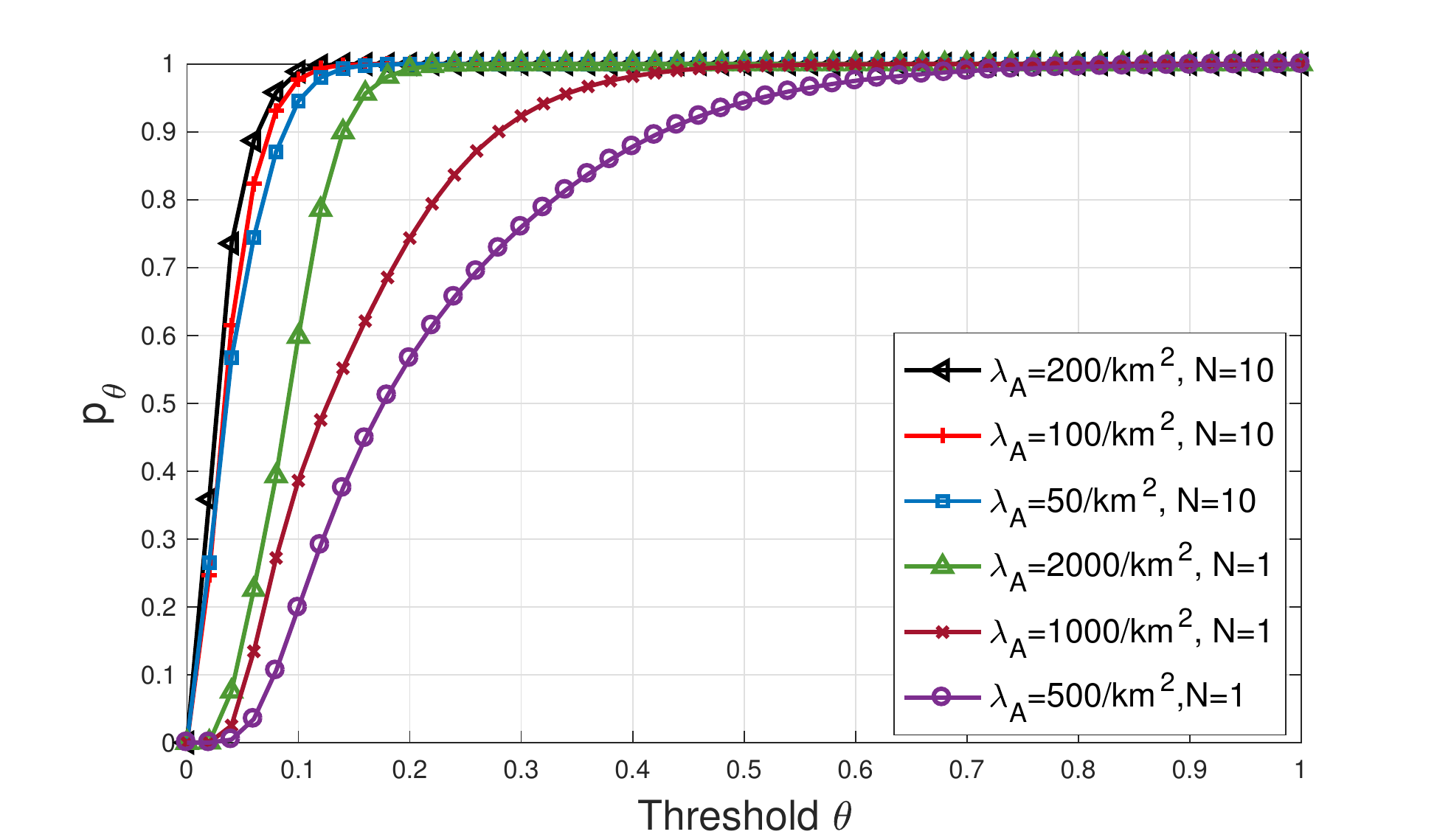}
	\caption{ The CDF of $X_{\text{ch}}$ with the three-slope pathloss model. $d_0=10$\,m and $d_1=50$\,m. The total antenna density is $\mu=N\lambda_A=\{500, 1000, 2000\}$/km$^2$.  }
	\label{fig:test_channel_three_slope}
\end{figure}
To validate our predictions, we present in Fig.~\ref{fig:test_channel_three_slope} $p_\theta$ (i.e., the CDF of $X_{\text{ch}}$) obtained with the three-slope pathloss model, with $d_0=$ 10\,m and $d_1=$ 50\,m. In this figure, we compare $p_\theta$ obtained with different values of the AP density $\lambda_A$ for $N=1$ and $N=10$ antennas per AP. First, with $N=1$, when the antenna density increases, the increase of $p_\theta$ is substantial, and more influential than the results in Fig.~\ref{fig:test_channel} and Fig.~\ref{fig:ptheta_diff_N_fixed_M} with the one-slope model. Second, comparing the results obtained with $N=1$ and $N=10$, we see that the number of antennas per AP plays a more important role than increasing the AP density in helping the channel to harden, in terms of achieving a small $X_{\text{ch}}$ with practically reasonable AP density values.

In summary, we have proved the following result.

\begin{theorem}
	With the three-slope pathloss model, due to the small pathloss exponent of the propagation environment nearby the user, the channel gain variance declines rather fast with the AP density compared to the mean value. Furthermore, having a large number of antennas per AP can guarantee small channel variation, which makes the channel hardening easier to achieve.  
\end{theorem}

\section{Favorable Propagation}
\label{sec:favorable}
In this section, we define and analyze the favorable propagation conditions in CF Massive MIMO networks. Similar to the previous section, we will consider both conventional CF networks with single-antenna APs and a generalization with multiple antennas per AP.

Recall that the channel vector from the $M$ antennas to the user $k$ is $\mathbf{g}_{k}=[g_{1,k},\ldots, g_{M,k}]^T$, where the $m$-th element is $g_{m,k}=\sqrt{l(d_{m,k})} h_{m,k}$.
To have favorable propagation, the channel vectors between the BS and the user terminals should be orthogonal, which means
\begin{equation}
\mathbf{g}_{k}^{H}\mathbf{g}_{j}= \left\{
\begin{array}{rcl}
&0 & \text{if} ~~~ k\neq j\\
&\|\mathbf{g}_{k}\|^2 \neq 0 & \text{if}  ~~~k=j.
\end{array} \right.
\end{equation}
When this condition is satisfied, each user can get the same communication performance as if it is alone in the network \cite{Rusek2013a}.
In practice, this condition is not fully satisfied, but can be approximately achieved when the number of antennas grows to infinity, in which case the channels are said to provide \textit{asymptotically favorable propagation}. To be more specific, in CF Massive MIMO, the asymptotically favorable propagation condition can be defined as follows:
\begin{equation}
\frac{\mathbf{g}_{k}^{H}\mathbf{g}_{j}}{\sqrt{\mathbb{E}\left[\|\mathbf{g}_{k}\|^2 | \mathbf{d}_k \right]\mathbb{E} \left[\|\mathbf{g}_{j}\|^2 | \mathbf{d}_j \right]}}\mathop{\longrightarrow }\limits 0, ~~~ \text{when}~~M\rightarrow \infty, k\neq j.
\label{eq:fav_propa_condition}
\end{equation}
Here, we have conditioned on a specific network realization with distance vectors $\mathbf{d}_k=[d_{1,k},\ldots,d_{M,k}]^T$ and $\mathbf{d}_j=[d_{1,j},\ldots,d_{M,j}]^T$, and each element $d_{m,k}$ represents the distance from the $m$-th antenna to the user $k$. Recall that every $N$ antennas are co-located at the same AP, we have $d_{(i-1)\cdot N+1,k}=\ldots=d_{i\cdot N,k}$ for $i=1,\ldots, L$.

Different from cellular Massive MIMO with co-located antennas, the large-scale fading coefficients from each antenna in CF Massive MIMO to a user are different, which can be viewed as a type of spatial channel correlation. Thus, we have
\begin{equation}
\mathbf{g}_{k}^{H}\mathbf{g}_{j}=\sum\limits_{m=1}^{M}\sqrt{l(d_{m,k}) l(d_{m,j})}h_{m,k}^{*}h_{m,j}
\label{eq:fav_propa_condition_nume}
\end{equation}
and
\begin{equation} \label{eq:fav_propa_condition_denom}
\sqrt{\mathbb{E} \left[\|\mathbf{g}_{k}\|^2 |\mathbf{d}_k\right]\mathbb{E} \left[\|\mathbf{g}_{j}\|^2 | \mathbf{d}_j\right]}=\sqrt{\sum\limits_{m=1}^{M}l(d_{m,k}) \sum\limits_{m=1}^{M}l(d_{m,j})}.
\end{equation}
Since $\{h_{m,k}\}$ are i.i.d. $\mathcal{CN}(0,1)$ RVs, we have
\begin{equation}
\mathbb{E}\left[ h_{m,k}^{*} h_{m,j}\right]= \left\{
\begin{array}{rcl}
0 & \text{if} & k\neq j\\
1 & \text{if} & k=j
\end{array} \right.
\end{equation}
and it follows that 
\begin{equation} 
\mathbb{E}\left[\frac{\mathbf{g}_{k}^{H}\mathbf{g}_{j}}{\sqrt{  \mathbb{E} \left[\|\mathbf{g}_{k}\|^2 | \mathbf{d}_k \right]\mathbb{E} \left[\|\mathbf{g}_{j}\|^2 | \mathbf{d}_j\right] }} \Bigg| \mathbf{d}_k, \mathbf{d}_j \right]= \left\{
\begin{array}{rcl}
&\!\!\!0 ~ &\text{if}  ~~k\neq j\\
&\!\!\!1 ~&\text{if} ~~ k=j.
\end{array} \right.
\label{eq:mean_fav_proba}
\end{equation}
With this mean value, the convergence in \eqref{eq:fav_propa_condition} holds (in mean square sense and in probability) if the variance of the left-hand side goes asymptotically to zero. Using \eqref{eq:fav_propa_condition_nume} and \eqref{eq:fav_propa_condition_denom}, we have
\begin{align}
&\text{Var}\left[\frac{\mathbf{g}_{k}^{H}\mathbf{g}_{j}}{\sqrt{  \mathbb{E} \left[\|\mathbf{g}_{k}\|^2 | \mathbf{d}_k \right]\mathbb{E} \left[\|\mathbf{g}_{j}\|^2 | \mathbf{d}_j\right] }} \Bigg| \mathbf{d}_k, \mathbf{d}_j \right] \nonumber \\=&\frac{N\sum\limits_{i=1}^{L} l(d_{i\cdot N,k})l(d_{i\cdot N,j})}{N^2 \sum\limits_{i=1}^{L}l(d_{i\cdot N,k})\sum\limits_{i=1}^{L}l(d_{i\cdot N,j})}  \label{eq:var_fav_proba} \\
\leq&\frac{L}{NL^2 \left(\frac{1}{L}\sum\limits_{i=1}^{L}l(d_{i\cdot N,k})\right)\left(\frac{1}{L}\sum\limits_{i=1}^{L}l(d_{i\cdot N,j})\right)}.
\label{eq:var_fav_proba_bound}
\end{align}
As the AP density $\lambda_A$ grows, $L$ increases. $\frac{1}{L}\sum\limits_{i=1}^{L}l(d_{i\cdot N,k})$ will approach $\mathbb{E}[l(d_{i\cdot N,k})]>0$, which is a positive value that only depends on the network region and pathloss model. Thus, \eqref{eq:var_fav_proba} is upper-bounded by a positive value that decreases as $1/L$ when $L$ increases. When $L\rightarrow \infty$, the variance of the channel orthogonality will approach $0$. Combined with \eqref{eq:mean_fav_proba}, we have proved that the asymptotically favorable propagation condition defined in \eqref{eq:fav_propa_condition} holds for CF Massive MIMO.

For finite $L$, we use \eqref{eq:var_fav_proba} to define the channel orthogonality metric 
\begin{equation}
X_{\text{fp}}=\frac{\sum\limits_{i=1}^{L}l(d_{i\cdot N,k})l(d_{i\cdot N,j})}{N\sum\limits_{i=1}^{L}l(d_{i\cdot N,k})\sum\limits_{i=1}^{L}l(d_{i\cdot N,j})}
\label{eq:fav_propa_NL}
\end{equation}
and consider the probability that two users at random locations have $X_{\text{fp}}$ no larger than a threshold $\gamma$:
\begin{equation}
p_{\gamma}=\mathbb{P}[X_{\text{fp}}\leq \gamma].
\end{equation}
Similar to the channel hardening measure defined in \eqref{eq:ptheta}, the probability is obtained over different network realizations that generate different distance vectors $\mathbf{d}_k$ and $\mathbf{d}_j$.
Clearly, in order to have asymptotically favorable propagation, when the antenna density grows, $p_{\gamma}$ should approach one for any $\gamma\geq0$. For practical purposes, it is desirable that $p_{\gamma}$ is large for values of the threshold $\gamma$ that are close to zero. In the following, we will analyze how the antenna density, the inter-user distance, and the pathloss exponent affects the channel orthogonality metric $X_{\text{fp}}$. For each case, we provide intuitive predications based on the analytical expression of $X_{\text{fp}}$ as a function of $\lambda_A$, which will be further validated by simulation results. 

\subsection{Impact of Antenna Density on the Channel Orthogonality}
\label{sec:impact_M}
The impact of the antenna density $\mu=N\lambda_A$ will be analyzed in two cases: fixed $\lambda_A$ with different $N$ or fixed $N$ with different $\lambda_A$. 
As mentioned above, $X_{\text{fp}}$ is inversely proportional to $N$ when $L$ is fixed, so increasing $N$ always helps the channels to become more orthogonal. In the other case, when $L$ increases, the denominator of $X_{\text{fp}}$ grows almost as $L^2$, while the numerator increases almost linearly with $L$. Since $L$ is a Poisson RV with mean value proportional to $\lambda_A$,
we have that $X_{\text{fp}}$ should scale roughly inversely proportional to $\lambda_A$, which evinces that for a given $\gamma$, $p_{\gamma}$ will grow with the AP density $\lambda_A$. Combining the two cases, we conclude that both larger $N$ and larger AP density $\lambda_A$ can help the channel to offer more favorable propagation. 

\begin{figure}[h!]
	\centering
	\includegraphics[width=0.5\textwidth]{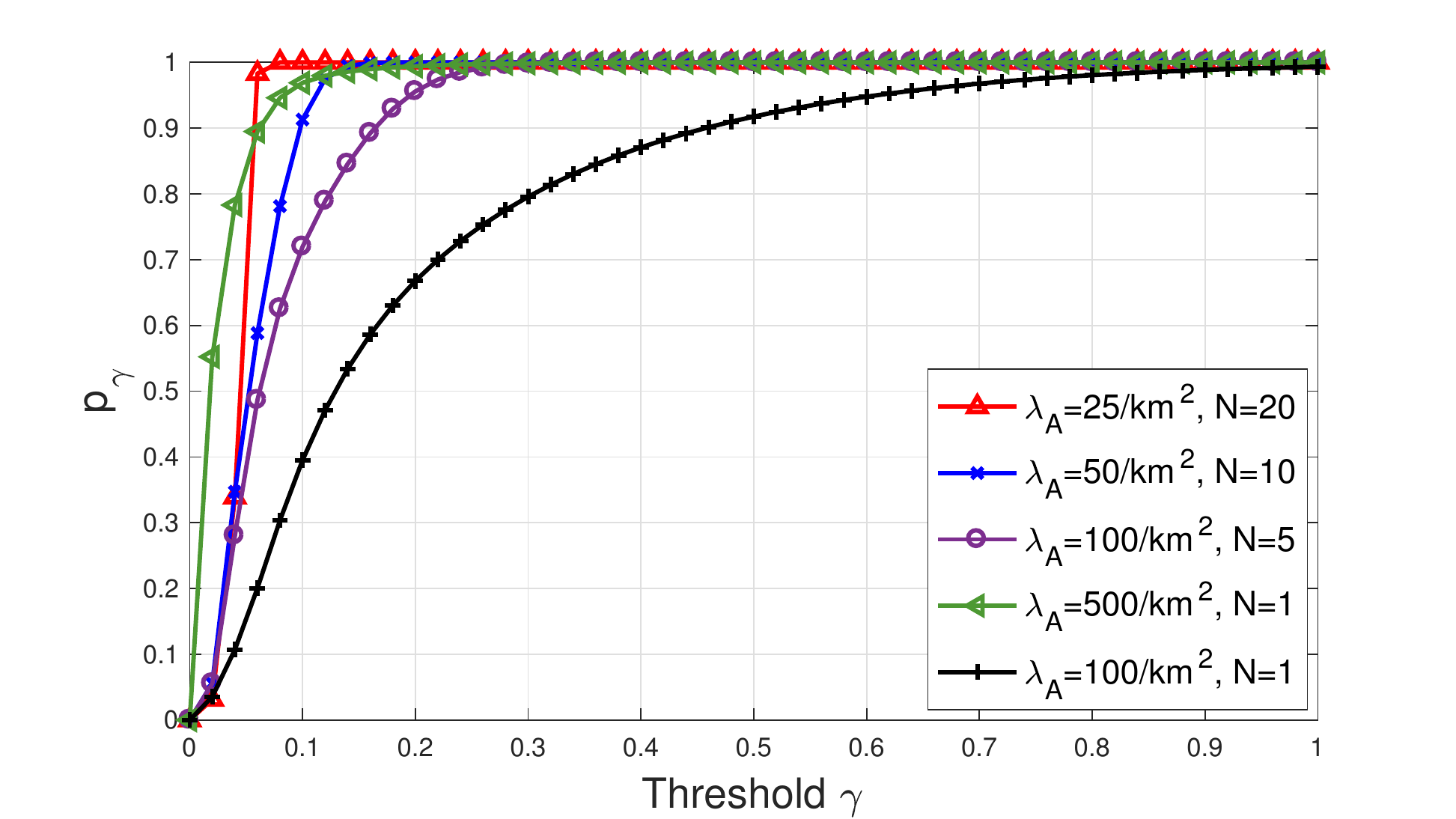}
	\caption{The CDF of $X_{\text{fp}}$. Pathloss exponent $\alpha=3.76$. Network radius $\rho=0.5$ km. For the first four curves (marked with upward triangle, cross, circle and left triangle), the antenna density is $\mu = N\lambda_A=500$/km$^2$ ($5\times10^{-4}$/m$^2$). The distance between user $j$ and user $k$ is $70$\,m.}
	\label{fig:test_fav_propa_diffN}
\end{figure}

Fig.~\ref{fig:test_fav_propa_diffN} presents the CDF of the channel orthogonality metric $X_{\text{fp}}$ with different $\lambda_A$ and $N$. By comparing the results obtained with $\lambda_A=\{500, 100\}$/km$^2$ and $N=\{1, 5\}$ (marked with circle, left triangle and plus sign), we validate our prediction that both increasing $\lambda_A$ and increasing $N$ can improve the channel orthogonality.

\begin{figure}[h!]
	\centering
	\includegraphics[width=0.5\textwidth]{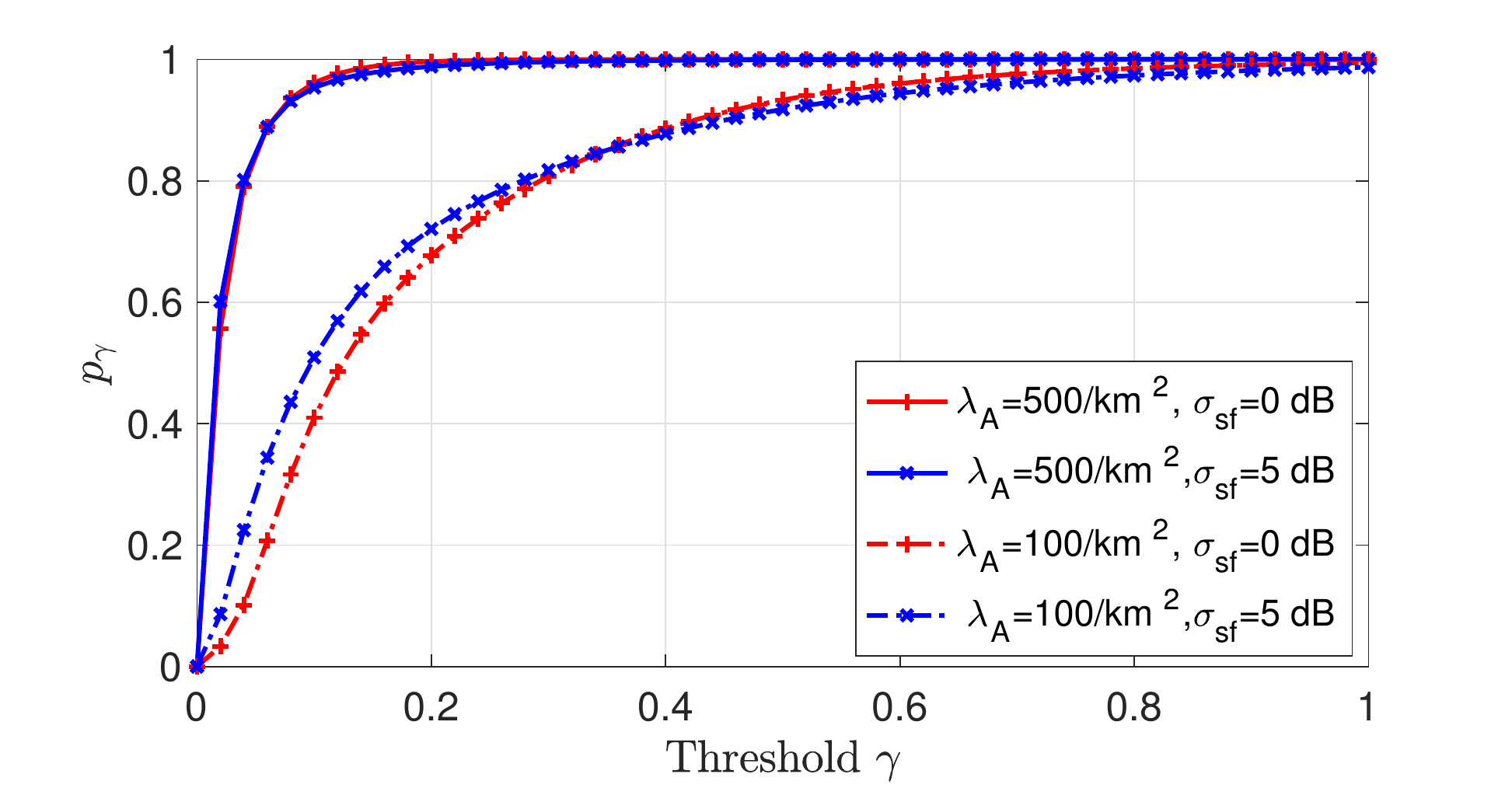}
	\caption{The CDF of $X'_{\text{fp}}$ with log-normal shadow fading. Same parameters as Fig.~\ref{fig:test_shadow}.} 
	\label{fig:test_shadow_fav_propa}
\end{figure}
 Similar to Section~\ref{sec:harden} , in the case with shadow fading, the large-scale fading from the $i$-th AP to the $k$-th user is $\beta_{i,k}=l(d_{i\cdot N,k})\cdot 10^{\frac{\sigma_{\text{sh}} z_{i,k}}{10}}$ with $z_{i,k}\sim \mathcal{N}(0,1)$ when $d_{i\cdot N,k}>50$\,m. The channel orthogonality metric in this case is defined by $X'_{\text{fp}}=\frac{\sum\limits_{i=1}^{L}\beta_{i,k} \beta_{i,j}}{N\sum\limits_{i=1}^{L}\beta_{i,k}\sum\limits_{i=1}^{L}\beta_{i,j}}$. In Fig.~\ref{fig:test_shadow_fav_propa}, we show the CDF of $X'_{\text{fp}}$ obtained by simulations. It is obvious that the inclusion of shadow fading does not affect much the distribution of $X'_{\text{fp}}$. Therefore, our conclusions on the impact of the antenna density on channel orthogonality will not be affected by whether shadow fading is included in the channel model.

\subsection{More Antennas on Few APs or More APs with Few Antennas?}
From \eqref{eq:fav_propa_NL}, we see that the value of $X_{\text{fp}}$ is always upper-bounded by $\frac{1}{N}$. When the antenna density $\mu=N\lambda_A$ is fixed, increasing $N$ means smaller $\lambda_A$. Hence, the average number of APs within close distance to the user will be less. Thus, it is hard to analytically predict whether it is more beneficial to have more antennas on few APs or more APs with few antennas.

In Fig.~\ref{fig:test_fav_propa_diffN}, we present the CDF of $X_{\text{fp}}$ when fixing the total antenna density $N\lambda_A=500$/km$^2$. We see that increasing $N$ does not necessarily lead to higher or lower $p_\gamma$ for a given value of $\gamma$. We also observe that when choosing sufficiently large $N$, e.g., $N\geq 20$,  the channel orthogonality metric $X_{\text{fp}}$ becomes very small. In other words, sufficiently large $N$ will help the channels to different users to be asymptotically orthogonal.

\subsection{Impact of Inter-User Distance on Channel Orthogonality}
From \eqref{eq:fav_propa_condition}, it is obvious that the distance between two users affects the variance of each term $\sqrt{l(d_{m,k}) l(d_{m,j})}h_{m,k}^{*}h_{m,j}$.  When two users are far apart, their channel vectors are more likely to be orthogonal with smaller variance. This result comes from the fact that $l(d_{m,k})l(d_{m,j})$ for all $m=1,\ldots, M$ will become much smaller when the distance $l_{k,j}$ between user $k$ and user $j$ is large, since there is no antenna in the network that is close to both users. In addition, $\sum_{m=1}^{M}l(d_{m,k}) \sum_{m=1}^{M}l(d_{m,j})$ will not vary much with the inter-user distance $l_{k,j}$ when $M$ is fixed. Therefore, $X_{\text{fp}}$ becomes smaller when the distance $l_{k,j}$ increases.

\begin{figure}[h!]
	\centering
	\includegraphics[width=0.5\textwidth]{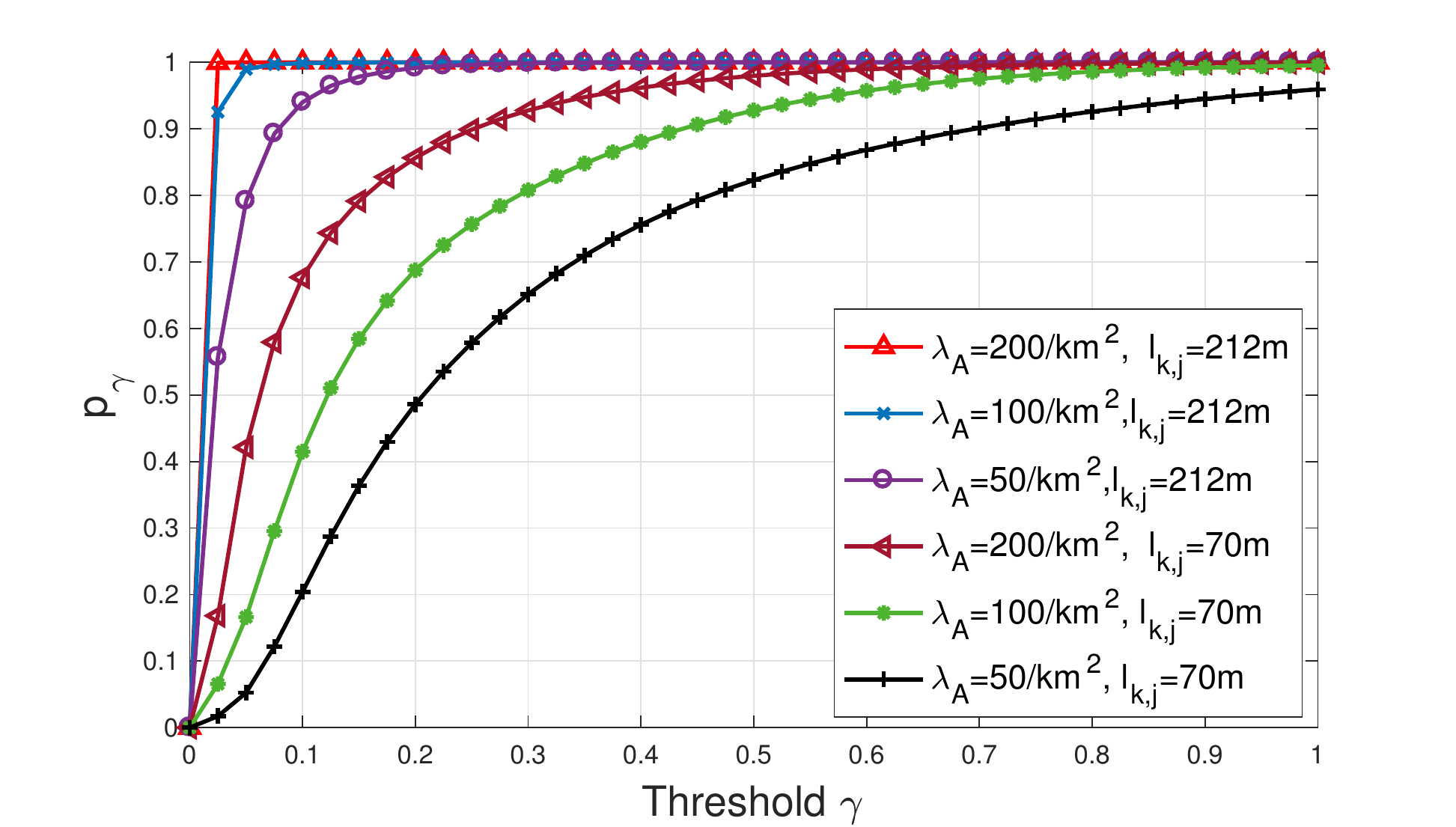}
	\caption{The CDF of $X_{\text{fp}}$ with different $\lambda_A$ and different inter-user distances $l_{d,j}$, $N=1$, and pathloss exponent $\alpha=3.76$.}
	\label{fig:test_fav_propa_diffM}
\end{figure}

In Fig.~\ref{fig:test_fav_propa_diffM}, we present $p_{\gamma}$ for different AP densities $\lambda_A \in \{50, \,100,\, 200\}$ APs/km$^2$, $N=1$,  and inter-user distances $l_{k,j}\in\{70, \,212\}$\,m. 
First, it is shown that with larger $\lambda_A$, the variance of the orthogonality metric $X_{\text{fp}}$ is smaller. 
Second, when the distance between two users is larger, they are more likely to have nearly orthogonal channels. This observation showcases the importance of serving spatially separated users in order to ensure near channel orthogonality. 

\subsection{Impact of Pathloss Exponent on the Channel Orthogonality}
When the number of antennas $M$ and their locations are fixed, we consider two extreme cases: user $k$ and user $j$ are very close or extremely far from each other.
Since $X_{\text{ch}}$ coincides with $X_{\text{fp}}$ in the special case when $d_{i\cdot N,k}\simeq d_{i\cdot N,j}$ for all $m=1,\ldots, M$, we infer from Section~\ref{sec:harden} that smaller pathloss exponent will also lead to smaller $X_{\text{fp}}$. In the other extreme case, when the two users are far apart, in the denominator of $X_{\text{fp}}$, the random realization of $\sum_{i=1}^{L}l(d_{i\cdot N,j})$ is almost independent of $\sum_{i=1}^{L}l(d_{i\cdot N,k})$, and both terms increase much faster than the numerator, especially when $\alpha$ is small. Combining these two extreme cases, we expect that smaller pathloss exponent would help the channels to become asymptotically orthogonal when $M$ is sufficiently large.

\begin{figure}[h!]
	\centering
	\includegraphics[width=0.5\textwidth]{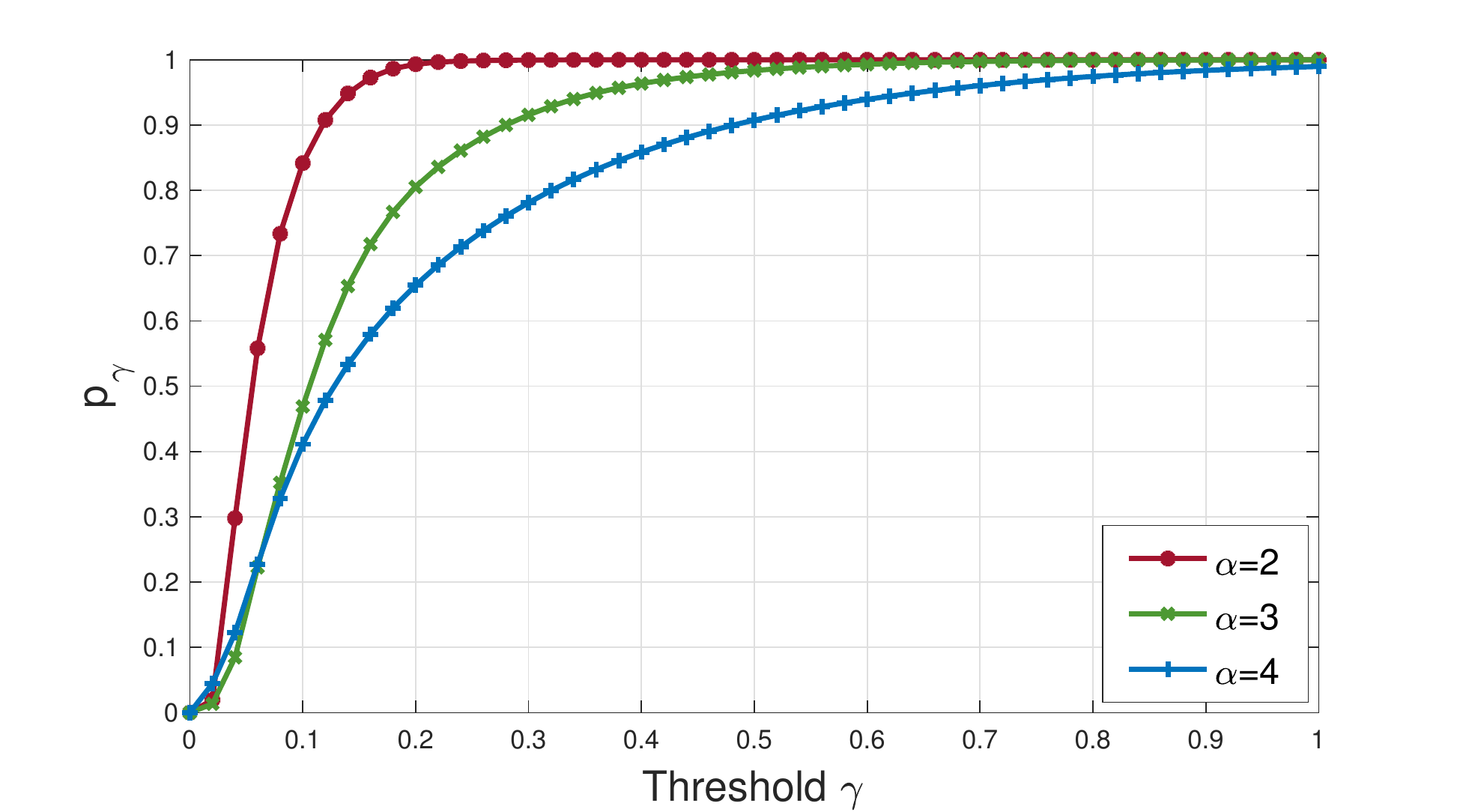}
	\caption{The CDF of $X_{\text{fp}}$ with different pathloss exponent $\alpha \in \{2, \,3,\, 4\}$, 
		$N=1$, and the inter-user distance $l_{k,j}=70$\,m.}
	\label{fig:test_fav_propa_alpha2}
\end{figure}

To confirm our prediction, in Fig.~\ref{fig:test_fav_propa_alpha2} we present $p_{\gamma}$ obtained with pathloss exponents $\alpha \in \{2, \,3,\, 4\}$, $N=1$, and an inter-user distance of $l_{k,j}=70$\,m. The figure shows that with smaller $\alpha$ the channels become more orthogonal, which is line with our prediction from above.  If we would instead use the three-slope pathloss model in \eqref{eq:three-slope}, APs close to the user (distance smaller than $d_1$) will have pathloss exponent $\alpha\leq 2$, and users at larger distances have pathloss exponent $3.5$. Therefore, compared to the single-slope non-singular pathloss model $l(r)=\min (r^{-3.76},1)$, the channels between two users will in the average be more orthogonal.

Summarizing the above analysis and observations, we have the following result.

\begin{theorem}
	Increasing the antenna density by increasing either the AP density or the number of antennas per AP can both help the user channels to offer favorable propagation. Smaller pathloss also helps the channels to become asymptotically orthogonal.
	The larger the distance between two users, the more likely their channels will be nearly orthogonal.
\end{theorem}

\section{Conclusions on Capacity Bounds for Cell-Free Massive MIMO} \label{sec:capacity-bounds}
A key conclusion from the previous sections is that CF Massive MIMO systems exhibit little channel hardening, as compared to cellular Massive MIMO. Hence, although CF Massive MIMO is mathematically equivalent to a single-cell Massive MIMO system with strong spatial channel correlation, we must be careful when reusing results from the Massive MIMO literature \cite{Huh2012a,Hoydis2013a,Yin2013a,SIG-093} on correlated channels. In particular, capacity bounds that were derived by relying on channel hardening can potentially be very loose when applied to CF systems. In this section, we review the standard capacity lower bounds and explain which ones are suitable for CF systems.

Consider a CF Massive MIMO system with $M$ single-antenna APs and $K$ users, which are assigned mutually orthogonal pilot sequences. TDD operation is assumed and the transmission is divided into coherence intervals of $\tau_c$ samples, whereof $\tau_p \leq \tau_c$ are used for uplink pilot signaling.
The channels are modeled as in previous sections: $\mathbf{g}_k \sim \mathcal{CN}(\mathbf{0},\mathbf{B}_k)$, where $\mathbf{B}_k = \mathrm{diag}(\beta_{1,k}, \ldots, \beta_{M,k})$ and  $\beta_{m,k}=l(r_{m,k})$.
Since we study the channel hardening and favorable propagation phenomena, and not the allocation of pilot sequences, the users are assumed to use mutually orthogonal pilot sequences (i.e., $\tau_p=K$).
However, as usual for correlated fading channels, if user $k$ and user $i$ have a small value of $\mathrm{tr}(\mathbf{B}_k \mathbf{B}_i)$, they can use the same pilot without causing much pilot contamination \cite[Sec.~4]{SIG-093}.\footnote{Such user pairs will also exhibit favorable propagation.}
User $k$ uses the transmit powers $\rho_k$ and $p_k$ for pilot and data, respectively.
Since the elements in $\mathbf{g}_k$ are independent, it is optimal to estimate them separately at the receiving antenna.
The MMSE estimate of $g_{m,k}$ is 
\begin{equation} \label{eq:MMSE-estimate}
\hat{g}_{m,k}=\frac{\sqrt{\tau_p \rho_k\beta_{m,k}}}{\tau_p \rho_k\beta_{m,k}+1}\left(\sqrt{\tau \rho_k}g_{m,k}+w_{m,k}\right),
\end{equation}
where $w_{m, k}\sim \mathcal{CN}(0,1)$ is i.i.d.~additive noise.
If we denote by $\gamma_{m,k}\mathop{=}\limits^{\triangle}\mathbb{E}\left[|\hat{g}_{m,k}|^2\right]$ the mean square of the MMSE estimate of $g_{m,k}$, then it follows from \eqref{eq:MMSE-estimate} that
\begin{equation} 
\gamma_{m,k}=\frac{\tau \rho_k\beta_{m,k}^2}{\tau \rho_k\beta_{m,k}+1}.
\end{equation}
We will now compare different achievable rate expressions for uplink and downlink when using MR processing, which is commonly assumed in CF Massive MIMO since it can be implemented distributively. The expressions are lower bounds on the ergodic capacity, thus we should use the one that gives the largest value to accurately predict the achievable performance.

The numerical part of the comparison considers a network area of $1$\,km $\times1$\,km. $K=20$ users are randomly and uniformly distributed in the network region, and we have $\tau_p=K$. The total number of antennas is $M=100$ in each simulation. All the APs are independently and uniformly distributed in the network region. The length of the coherence block is $\tau_c=500$. The large-scale fading coefficients between the antennas and the users are generated from $300$ different PPP realizations. The three-slope pathloss model in \eqref{eq:three-slope} is used with $d_0=10$\,m and $d_1=50$\,m. The constant factor $C$ (dB) is given by 
\begin{equation}
\begin{split}
C=&105+94-46.3-33.9\log_{10}(f)+13.82\log_{10}(h_{AP}) \\&+ (1.1\log_{10}(f)-0.7)h_u - (1.56\log_{10}(f)-0.8),
\end{split}
\end{equation}
where $f$ is the carrier frequency, $h_{AP}$ and $h_{u}$ are the AP and user antenna height, respectively \cite{CF_mimo}. Here, $105 = 35 \log_{10}(10^3)$ comes from the fact that the pathloss model in \cite{CF_mimo} is measured in kilometer instead of meter, and $94$ comes from the noise variance. The transmit power per user is 100\,mW.
These simulation parameters are the same as in \cite{CF_mimo} and are summarized in Table~\ref{system_params}.
Note that we are considering fixed power values and antenna densities since, in this section, we want to evaluate the accuracy of difference capacity bounds in a practical setup.

\begin{figure}[t!]
	\centering
		\includegraphics[width=0.5\textwidth]{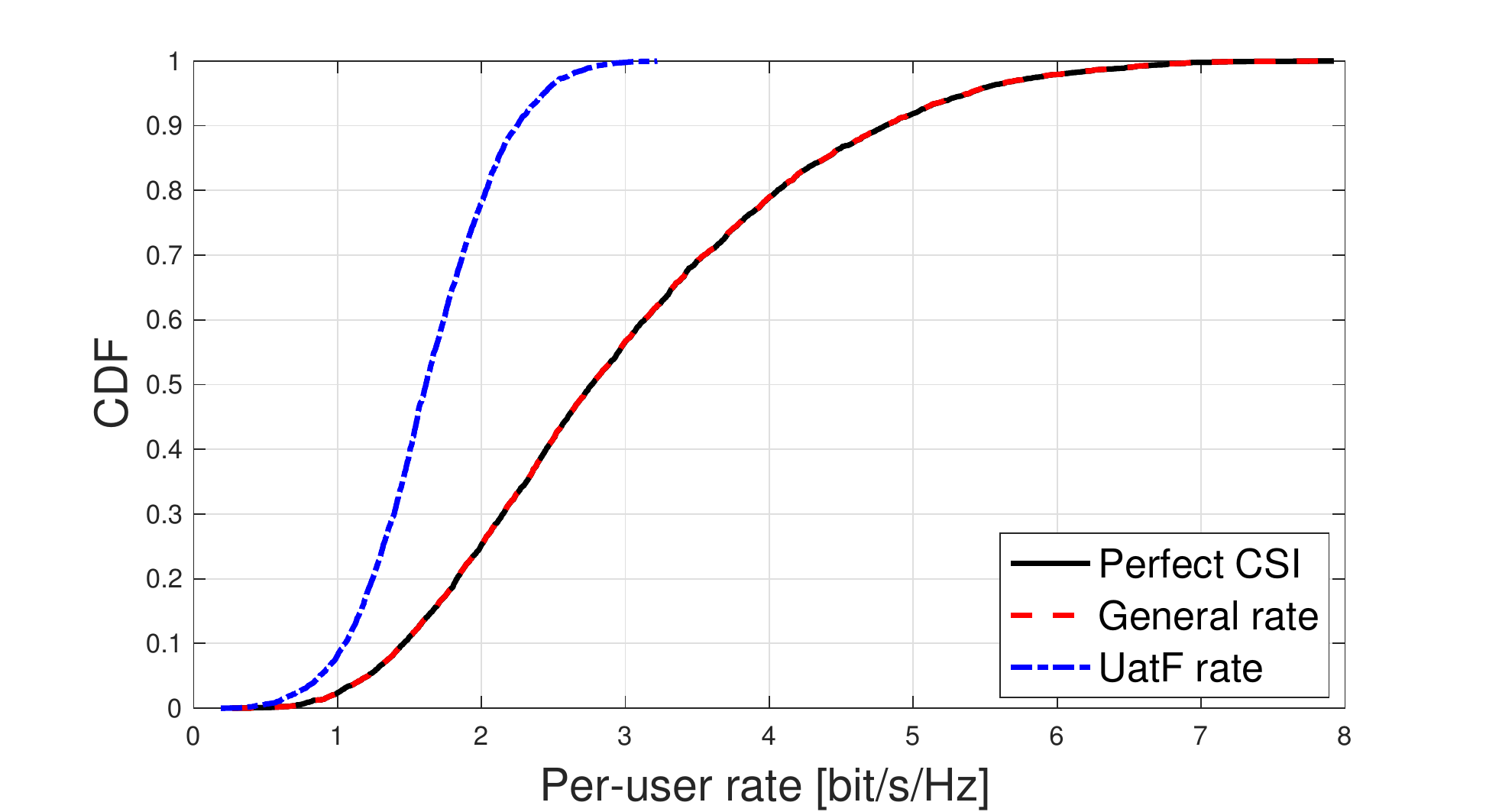}
		\caption{The CDF of the uplink achievable rates obtained with the UatF bound, the general bound, and the rate with perfect CSI. $M=100$, $N=1$, $K=20$, $\tau_p=20$.}
		\label{fig:UL-bound}
\end{figure}
\begin{figure}
	\centering
		\includegraphics[width=0.5\textwidth]{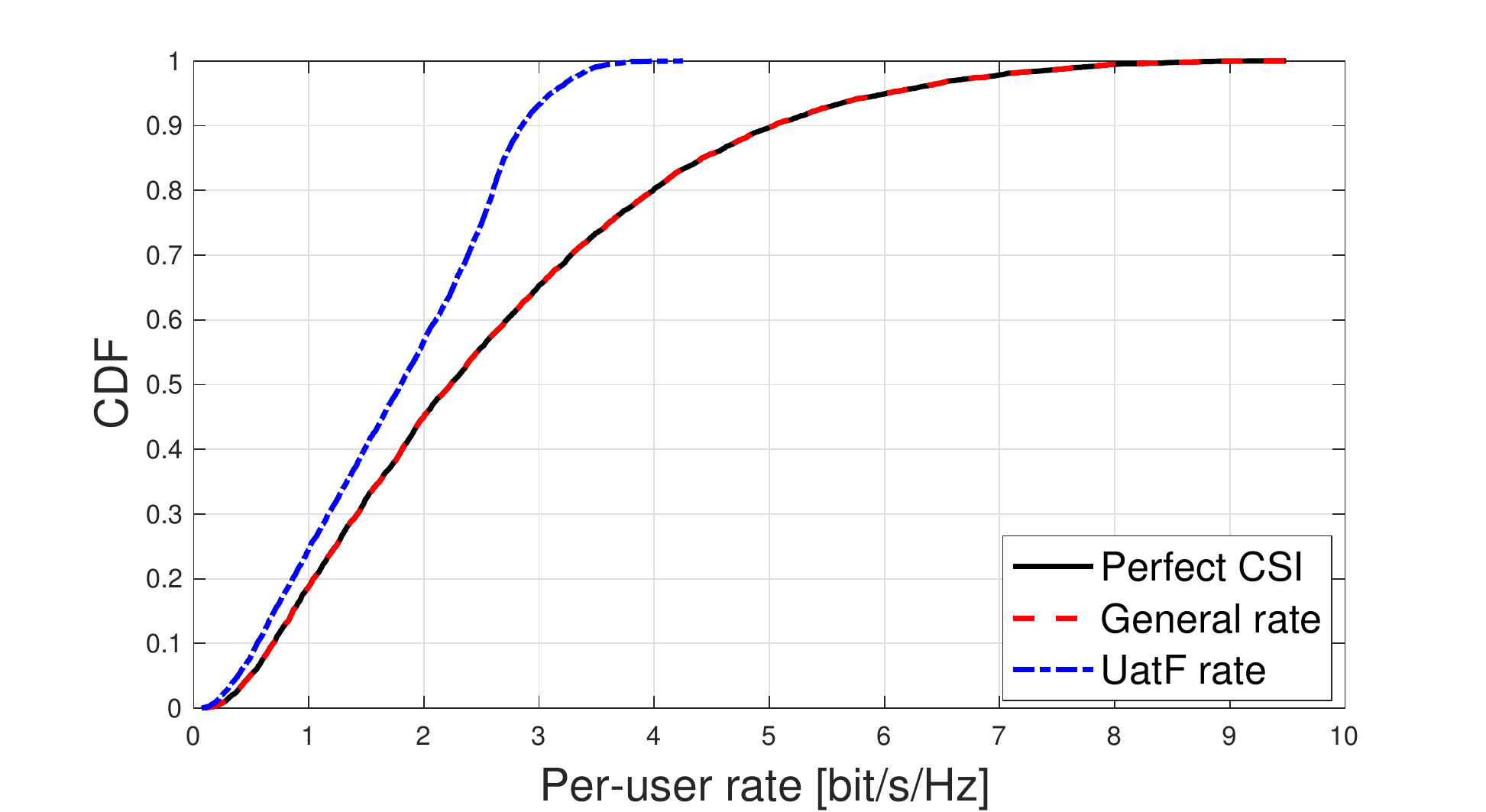}
		\caption{Same as Fig.~\ref{fig:UL-bound}. $M=100$, $N=5$, $K=20$, $\tau_p=20$.}
		\label{fig:UL-bound-N5}
\end{figure}

\begin{table}[ht!]
	\centering
	\caption{Simulation Setup}
	\renewcommand{\arraystretch}{1.1}
	\begin{tabular}{|c|c|}
		\firsthline
		\textbf{Parameters}              & \textbf{Values}  \\
		\hline   
		\small {Number of antennas: $M$}& \small {$100$}    \\
		\small {Number of antennas per AP: $N$}& \small {Varying}    \\
		\small {Number of users: $K$}& \small {$20$}    \\
		\small {Pilot length: $\tau_p$}& \small {$20$}    \\	
		\small {Length of the coherence block: $\tau_c$}& \small {$500$ }    \\
		\small {Distances in three-slope model: $d_0$, $d_1$    } & \small {$10$ m, $50$ m  }    \\
		\small {Carrier frequency: $f$ }    & \small { $1.9$ GHz}   \\  
		\small {Antenna heights: $h_{AP}$, $h_u$ }    & \small { $15$ m, $1.65$ m}   \\  
		\small {Uplink pilot power $\rho_{k}$}  & \small {$100$ mW} \\
		\small {Uplink data power $p_{k}$ }  & \small {$100$ mW} \\
		\small {Downlink power per user: $q$}    & \small {$100$ mW}   \\   
		\lasthline
	\end{tabular}
	\label{system_params}
\end{table}

\subsection{Uplink Achievable Rate}
\label{sec:UL-bound}
During the uplink data transmission, all $K$ users simultaneously transmit to the $M$ APs. When using MR, an achievable rate (i.e., a lower bound on the capacity) of user $k$ is
\begin{equation}
R_{\text{u},k}^{\textrm{UatF}}=   \log_{2}\left(1+\frac{p_{k} \left(\sum\limits_{m=1}^{M}\gamma_{m,k} \right)^2}{ \sum\limits_{j=1}^{K}p_{j}\sum\limits_{m=1}^{M}\gamma_{m,k} \beta_{m,j}+\sum\limits_{m=1}^{M}\gamma_{m,k}}\right), \label{eq:bound-uplink}
\end{equation}
which was used for CF Massive MIMO in \cite{CF_mimo}. This bound is derived based on the use and then forget (UatF) principle \cite{Marzetta2016a}, where the channel estimates are used for MR but then ``forgotten'' and channel hardening is utilized to obtain a simple closed-form expression. Note that $R_{\text{u},k}^{\textrm{UatF}}$ is a special case of the general expression in \cite{Bjornson2016c} for correlated single-cell Massive MIMO systems that apply MR processing.
Alternatively, the achievable rate expression in \cite{Hoydis2013a} for spatially correlated channels can be used:
\begin{equation}
\begin{split}
&R_{\text{u},k}= \\&\mathbb{E}\left[\log_{2}\left(1+\frac{p_k |\mathbf{a}_{k}^{H}\hat{\mathbf{g}}_{k}|^2}{\sum\limits_{j\neq k}^{K} p_{j} |\mathbf{a}_{k}^{H}\hat{\mathbf{g}}_{j}|^2+ \mathbf{a}_{k}^{H} \Big( \sum\limits_{j=1}^{K}p_{j}(\mathbf{B}_{j}-\boldsymbol{\Gamma}_{j}) + \mathbf{I}_M \Big) \mathbf{a}_{k} }\right)\right],
\end{split}
\label{eq:bound-general}
\end{equation}
where $\hat{\mathbf{g}}_{k}=[\hat{g}_{1,k},\ldots, \hat{g}_{M,k}]^T$, $\boldsymbol{\Gamma}_{j} = \mathrm{diag}(\gamma_{1,j}, \ldots, \gamma_{M,j})$, and $\mathbf{a}_{k}$ is the combining vector, which is $\mathbf{a}_{k}=\hat{\mathbf{g}}_{k}$ for MR. This general bound does not rely on channel hardening.

In Fig.~\ref{fig:UL-bound} and Fig.~\ref{fig:UL-bound-N5}, we compare the uplink achievable rates obtained with \eqref{eq:bound-uplink} and \eqref{eq:bound-general}. As a reference, we also provide the rate with perfect CSI (obtained from \eqref{eq:bound-general} by letting $\rho_k \to \infty$). 
Fig.~\ref{fig:UL-bound} and Fig.~\ref{fig:UL-bound-N5} show the CDFs of the user rates for random distances between the $L$ APs and the $K$ users.
To compare the impact of different number of APs, the results in Fig.~\ref{fig:UL-bound} are obtained with $L=100$ single-antenna APs and those in Fig.~\ref{fig:UL-bound-N5} are obtained with $L=20$ APs with $N=5$ antennas per AP.
	
The general rate expression in \eqref{eq:bound-general} provides almost the same rates as the perfect CSI case, which indicates that the estimation errors to the closest APs are negligible and these are the ones that have a non-negligible impact on the rate.
In contrast, the UatF rate in \eqref{eq:bound-uplink} is a much looser capacity bound, particularly for the users that support the highest data rates; some users get almost twice the rate when using the general rate expression. The gap is due to the lack of channel hardening, because the UatF rate require channel hardening to be a tight bound.
Hence, a general guideline is to only use \eqref{eq:bound-general} when evaluating the achievable rates in CF Massive MIMO with single-antenna APs. This guideline applies also to non-distributed processing schemes, such as regularized zero-forcing, which are used to cancel interference at the cost of further reducing the channel hardening effect \cite[Sec.~4.1.6]{SIG-093}.

When comparing Fig.~\ref{fig:UL-bound} and Fig.~\ref{fig:UL-bound-N5}, we note that 
the difference between the UatF rate and the perfect CSI case reduces when having multiple antennas per AP (e.g., $N=5$). The reason is that having multiple antennas per AP leads to more channel hardening, as we have shown in Section~\ref{sec:harden}. However, the average rate reduces when going from many single-antenna APs to fewer multi-antenna APs, due to the loss in macro-diversity.
If providing high data rates is the first priority, then it is always better to deploy as many single-antenna APs as possible. In this case, the achievable rates should be obtained without the channel hardening assumption; in fact, the general bound in \eqref{eq:bound-general} is always preferable.

\subsection{Downlink Achievable Rate}
\label{sec:DL-bound}

The user decodes the received signals in the downlink. It is generally suboptimal to use explicit downlink pilots for channel estimation at the user \cite{channel_hardening}, but the optimal detection scheme is still unknown.
In Massive MIMO, it has been a common practice to use rate expressions where the detector relies on channel hardening, by presuming that the instantaneous precoded channel is close to its mean value. If MR precoding is used, then one can use the rate expression for spatially correlated fading from \cite{Hoydis2013a} and obtain the achievable rate
\begin{equation}
R_{\text{d},k}^{\textrm{UatF}}=\log_{2}\left(1+\frac{\left(\sum\limits_{m=1}^{M} \sqrt{p_{m,k} \gamma_{m,k}} \right)^2}{ \sum\limits_{k'=1}^{K}\sum\limits_{m=1}^{M}p_{m,k'}\beta_{m,k}+1}\right) \label{eq:bound-downlink}
\end{equation}
for user $k$, where $p_{m,k}$ is the average transmit power allocated to user $k$ by AP $m$. This type of expression was used for CF Massive MIMO in \cite{CF_mimo}, using a slightly different notation. We call this the UatF rate since we use the received signals for detection, but then ``forget'' to use them for blind estimation of the instantaneous channel realizations. Note that $R_{\text{d},k}^{\textrm{UatF}}$ is always a rigorous lower bound on the capacity, but this does not mean that the bound is tight.

Suppose $\mathbf{a}_{k}$ is the precoding vector (including power allocation) assigned to user $k$. To avoid relying on channel hardening, user $k$ can estimate its instantaneous channel after precoding, $\mathbf{a}_{k}^{H} \mathbf{g}_{k}$, from the collection of  $\tau_d$ received downlink signals in the current coherence block. Note that no explicit downlink pilots are needed for this task \cite{Caire2017a,SIG-093}, since only a scalar needs to be deduced from the received signals. By following the rigorous capacity bounding technique in \cite[Lemma 3]{Caire2017a}, we obtain the achievable rate
\begin{equation} 
\begin{split}
R_{\text{d},k}=&\mathbb{E}\Bigg[\log_{2}\Bigg(1+\frac{|\mathbf{a}_{k}^{H} \mathbf{g}_{k}|^2}{\sum\limits_{j\neq k}^{K} |\mathbf{a}_{j}^{H}\mathbf{g}_{k}|^2+ 1 }\Bigg)\Bigg]\\ &- \frac{1}{\tau_d} \sum_{j=1}^{K} \log_2 \left( 1 + \tau_d\text{Var} [\mathbf{a}_{j}^{H}\mathbf{g}_{k}]   \right).
\end{split}
\label{eq:downlink-general}
\end{equation}
The first term in \eqref{eq:downlink-general} is the rate with perfect CSI and the second term is a penalty term from imperfect channel estimation at the user. Note that the latter term vanishes as $\tau_d \to \infty$, thus this general rate expression is a good lower bound when the channels change slowly. The rate with MR precoding is obtained by setting $\mathbf{a}_{k} = \big[  \hat{g}_{1,k} \sqrt{\frac{p_{1,k}}{\gamma_{1,k}} } \, \ldots \, \hat{g}_{M,k} \sqrt{\frac{p_{M,k}}{\gamma_{M,k}} } \big]^T$ in \eqref{eq:downlink-general}.

\begin{figure}[h!]
	\centering
	\includegraphics[width=0.5\textwidth]{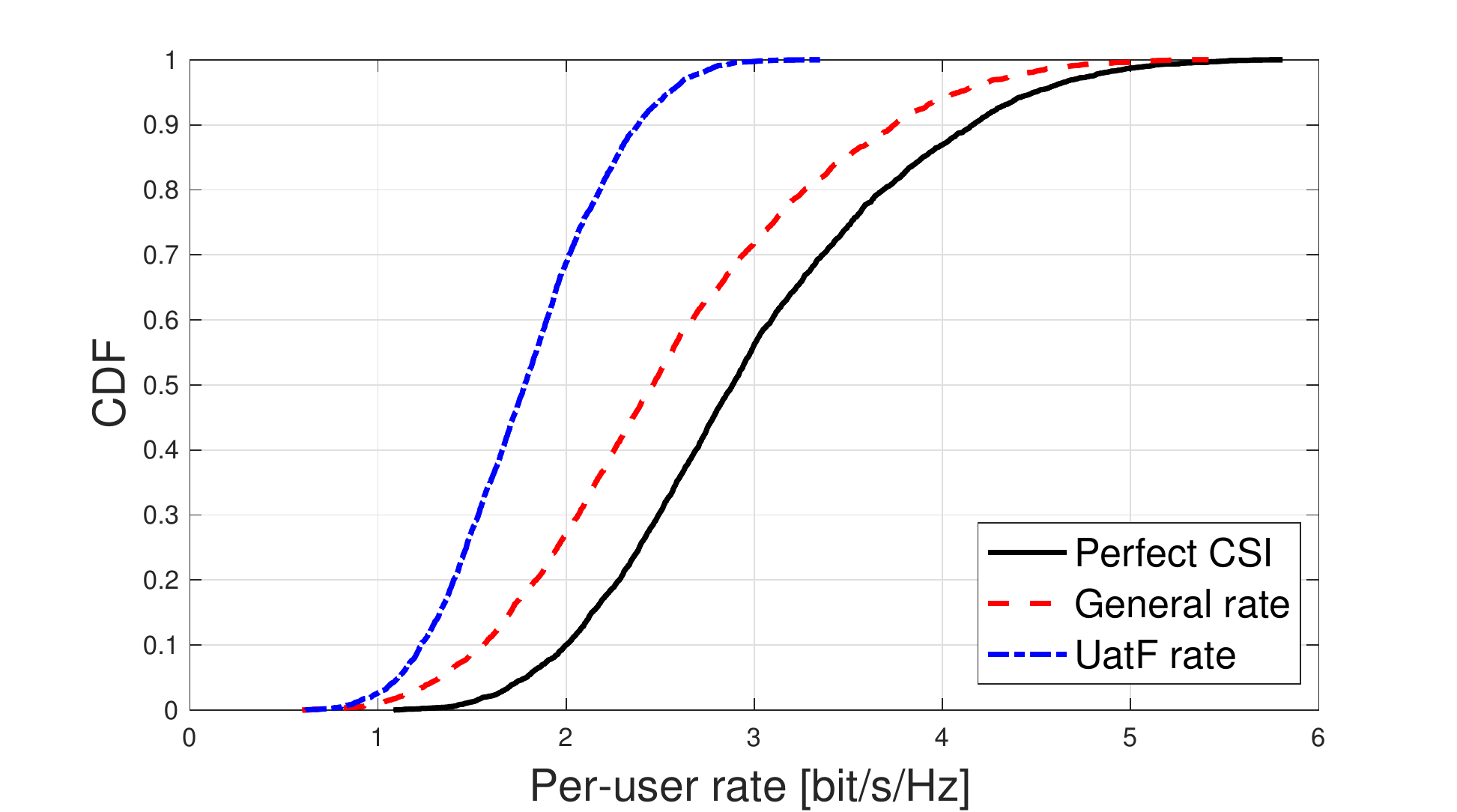}
	\caption{The CDF of the downlink achievable rates obtained with the UatF bound, the general bound, and the rate with perfect CSI. $M=100$, $N=1$, $K=20$, $\tau_p=20$.}
	\label{fig:DL-bound}
\end{figure}

In Fig.~\ref{fig:DL-bound}, we compare the downlink achievable rates obtained with \eqref{eq:bound-downlink} and \eqref{eq:downlink-general}. We also consider the rate with perfect CSI at each user (obtained from \eqref{eq:downlink-general} by letting $\tau_d \to \infty$). The simulation parameters are the same as in the uplink. The downlink power $p_{m,k}$ is given by
\begin{equation}
p_{m,k}=\frac{q\cdot \gamma_{m,k}}{\mathbb{E}[\|\hat{\mathbf{g}}_k\|^2]}=q\frac{\gamma_{m,k}}{\sum\limits_{m'=1}^{M}\gamma_{m',k}},
\end{equation}
where $q$ is the downlink power allocated to each user $k$, which is chosen as $100$ mW in the simulations. By doing so, we have  $\mathbb{E}[\|\mathbf{a}_k\|^2]=q$, which is the same for all users.

Fig.~\ref{fig:DL-bound} shows the CDFs of the user rates for random antenna-user distances. As in the uplink, there is a substantial gap between the rates achieved in the perfect CSI case and the UatF rate.
The curve for the general rate in \eqref{eq:downlink-general} is in the middle, which implies that the users should (somehow) estimate their instantaneous downlink channels and not only rely on channel hardening in CF Massive MIMO. While the gap to the perfect CSI curve vanishes as $\tau_d \to \infty$, it is unknown if the gap for the considered $\tau_d = \tau_c-\tau_p = 480$ is due to a fundamentally limited estimation quality or an artifact from the capacity bounding technique in \cite[Lemma 3]{Caire2017a} (i.e., the penalty term might be substantially larger than it should). In any case, there is a need to further study the achievable downlink rates in CF Massive MIMO.

\section{Conclusion}
\label{sec:conclusion}
In this work, we provided a thorough investigation of the channel hardening and favorable propagation phenomena in CF Massive MIMO systems from a stochastic geometry perspective. By studying the channel distribution from stochastically distributed APs with either a single antenna or multiple antennas per AP, we characterized the channel gain distribution. Based on this result, we examined the conditions for when channel hardening and favorable propagation occur. Our results show that whether or not the channel hardens as the number of APs increases depends strongly on the propagation environment and pathloss model. In general, one should not expect much hardening.
However, one can obtain more hardening by deploying multiple antennas per AP; for a given antenna density, it is beneficial to have a few multi-antenna APs than many single-antenna APs. There are several factors that can help the channel to provide favorable propagation, such as a smaller pathloss exponent, higher antenna density, and spatially separated users with larger distances. Spatially well-separated users will generally exhibit favorable propagation since they are essentially communicating with different subsets of the APs.  For a given antenna density, having multi-antenna APs is not necessarily better than having many single-antenna APs in providing favorable propagation, but it depends on the propagation scenario.

One main implication of this work is that one should not rely on channel hardening and favorable propagation when computing the achievable rates in CF Massive MIMO, because this could lead to a great underestimation of the achievable performance. There is a good uplink rate expression for spatially correlated Massive MIMO systems that can be used. Further development of downlink rate expressions is needed to fully understand the achievable downlink performance in CF Massive MIMO.

\appendix
\appendices
\subsection{Proof of Lemma~\ref{lemma1}}
\label{appen1}
Recall Campbell's theorem as follows \cite{haenggi2009interference}. If $f(x):\mathbb{R}^{d}\rightarrow[0,+\infty]$ is a measurable function and $\Phi$ is a stationary/homogeneous PPP with density $\lambda$, then
\begin{equation}
\mathbb{E}\left[\sum\limits_{x\in \Phi}f(x)\right]=\lambda\int_{\mathbb{R}^{d}} f(x) \text{d}x.
\end{equation}
Since $\Phi_A$ is a two-dimensional homogeneous PPP, for a finite network region with radius $\rho$, we have 
\begin{align}
\mathbb{E}\left[\|\mathbf{g}_k\|^2\right]&= \mathbb{E}\left[\sum\limits_{i\in \Phi_A}  H_{i} l(r_{i})\right] = \mathbb{E}\left[H_{i}\right]\cdot \lambda_A\int_{\mathcal{B}(0, \rho)} l(\|x\|)  \text{d}x
\nonumber\\&=\lambda_A\cdot \mathbb{E}\left[H_{i}\right]2\pi\int_{0}^{\rho} l(r) r\text{d}r.
\end{align}
Note that $l(r)=\min(1,r^{-\alpha})$ and $\mathbb{E}\left[H_{i}\right]=N$ as a result of the Gamma distribution, thus
\begin{equation}
\mathbb{E}\left[\|\mathbf{g}_k\|^2\right]=N\lambda_A2\pi\left(\int_{0}^{1} r\text{d}r+\int_{1}^{\rho}  r^{1-\alpha}\text{d}r\right).
\end{equation}
Depending on the value of $\alpha$, we have
\begin{align}
\mathbb{E}\left[\|\mathbf{g}_k\|^2\right]= \left\{
\begin{array}{rcl}
&\!\!\!\!N\lambda_A\pi\left(1+\frac{2\left(1-\rho^{2-\alpha}\right) }{\alpha-2}\right) & \text{if}~~\alpha\neq 2\\
&\!\!\!\!N\lambda_A\pi\left(1+2\ln \rho \right)& \text{if}~~\alpha=2\\
\end{array} \right.
\end{align}
From \cite{haenggi2009interference}, we have the expression for the variance 
\begin{equation}
\text{Var} \left[\sum\limits_{x\in \Phi}f(x)\right]=\lambda\int_{\mathbb{R}^{d}} f(x)^2 \text{d}x.
\end{equation}
Since $H_{i}\sim\text{Gamma}(N,1)$, we have $\mathbb{E}\left[H_{i}^2\right]=\left(\mathbb{E}\left[H_{i}\right]\right)^2+\text{Var}\left[H_{i}^2\right]=N^2+N$, thus 
\begin{align}
\text{Var}\left[\|\mathbf{g}_k\|^2\right]
&=\lambda_A\cdot \mathbb{E}\left[H_{i}^2\right]2\pi\int_{0}^{\rho} l^2(r) r\text{d}r \nonumber\\
&=(N^2+N)\lambda_A2\pi\left(\int_{0}^{1} r\text{d}r+\int_{1}^{\infty}  r^{1-2\alpha}\text{d}r\right) \nonumber\\
&=(N^2+N)\lambda_A\pi\left(1+\frac{1}{\alpha-1}\left(1-\rho^{2-2\alpha}\right)\right).
\label{eq:var_alpha}
\end{align}
Note that \eqref{eq:var_alpha} will have a different form when $\alpha=1$, which is unlikely to happen in a real propagation environment. Thus, this case is not discussed here.

\subsection{Proof of Lemma~\ref{lemma2}}
\label{appen2}
Using Campbell's theorem 
\begin{align}
\mathbb{E}\left[Y_1\right]= &\mathbb{E}\left[\sum_{i=1}^{L}  l(r_{i})\right] \nonumber\\
=  &\lambda_A\int_{\mathcal{B}(0, \rho)} l(\|x\|)  \text{d}x=\lambda_A\cdot2\pi\int_{0}^{\rho} l(r) r\text{d}r.
\end{align}
With $l(r)=\min(1,r^{-\alpha})$, following the same steps as in Appendix~\ref{appen1}, we obtain \eqref{eq:mean_Y1}.
Similarly, we have the variance of $Y_1$ as
\begin{align}
\text{Var}\left[Y_1\right]
=&\lambda_A2\pi\int_{0}^{\rho} l^2(r) r\text{d}r\nonumber\\
=&\lambda_A\pi\left(1+\frac{1}{\alpha-1}\left(1-\rho^{2-2\alpha}\right)\right).
\end{align}

\begin{IEEEbiography}[{\includegraphics[width=1in,height=1.25in,clip,keepaspectratio]{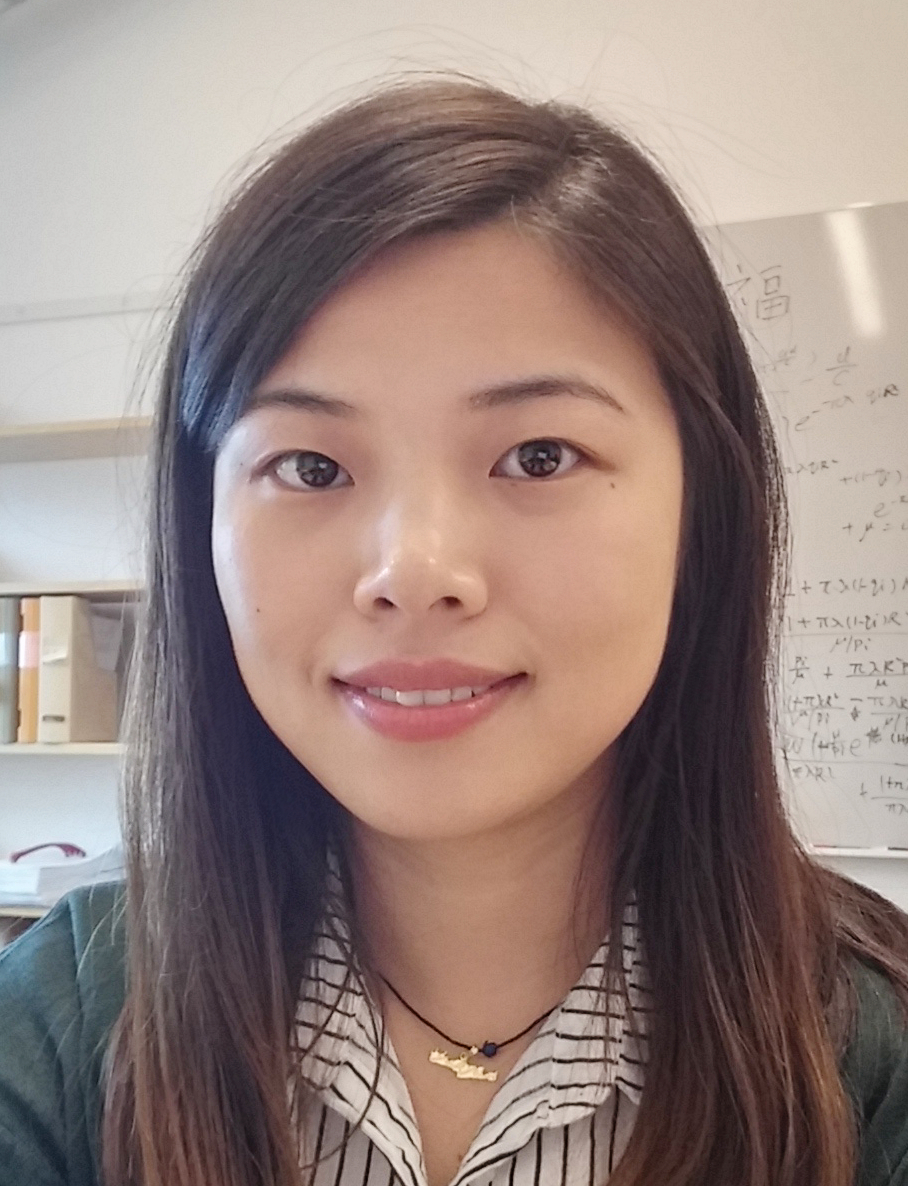}}]
	{\bfseries Zheng Chen} (S'14-M'17) received the B.S. degree from Huazhong University of Science and Technology (HUST), Wuhan, China, in 2011. She received the M.S. and Ph.D. degrees from CentraleSup\'{e}lec, Gif-sur-Yvette, France, in 2013 and 2016, respectively. From June to November 2015, she was a visiting scholar at Singapore University of Technology and Design (SUTD), Singapore. Since January 2017, she has been a Postdoctoral Researcher in Link\"{o}ping University, Link\"{o}ping, Sweden. Her research interests include stochastic geometry, queueing analysis, stochastic optimization, device-to-device communication, and wireless caching networks. She was selected as an Exemplary Reviewer for IEEE Communications Letters in 2016 and the Best Reviewer for IEEE Transactions on Wireless Communications in 2017.
\end{IEEEbiography}

\vspace{-10cm}
\begin{IEEEbiography}[{\includegraphics[width=1in,height=1.25in,keepaspectratio]{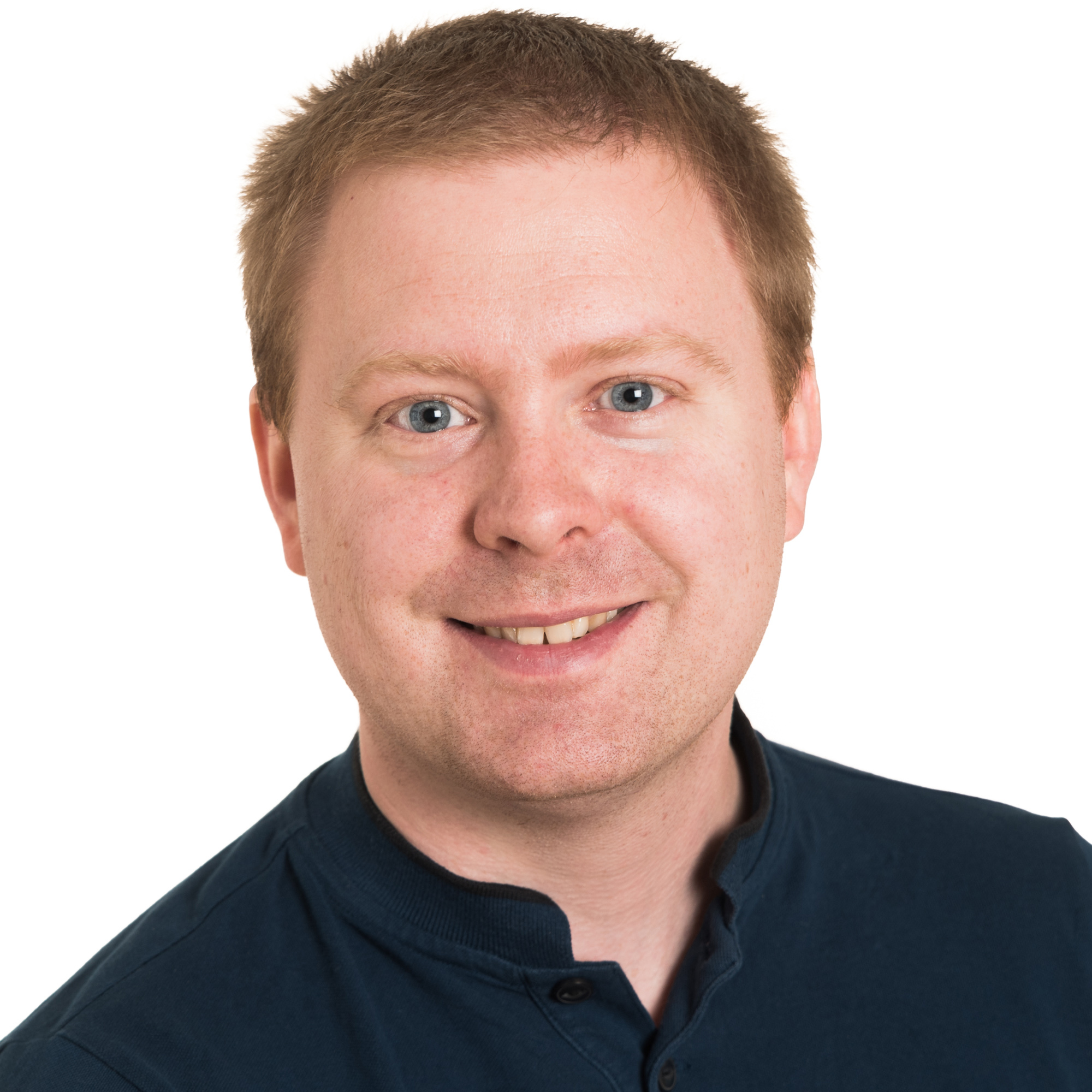}}]
	{\bfseries Emil Bj\"ornson}  (S'07-M'12-SM'17) received the M.S. degree in Engineering Mathematics from Lund University, Sweden, in 2007. He received the Ph.D. degree in Telecommunications from KTH Royal Institute of Technology, Sweden, in 2011. From 2012 to mid 2014, he was a joint postdoc at the Alcatel-Lucent Chair on Flexible Radio, SUPELEC, France, and at KTH. He joined Link\"oping University, Sweden, in 2014 and is currently Associate Professor and Docent at the Division of Communication Systems.
	
	He performs research on multi-antenna communications, Massive MIMO, radio resource allocation, energy-efficient communications, and network design. He is on the editorial board of the IEEE Transactions on Communications (since 2017) and the IEEE Transactions on Green Communications and Networking (since 2016). He is the first author of the textbooks ``Massive MIMO Networks: Spectral, Energy, and Hardware Efficiency'' (2017)  and ``Optimal Resource Allocation in Coordinated Multi-Cell Systems'' from 2013. He is dedicated to reproducible research and has made a large amount of simulation code publicly available.
	
	Dr. Bj\"ornson has performed MIMO research for more than ten years and has filed more than ten related patent applications. He received the 2018 Marconi Prize Paper Award in Wireless Communications, the 2016 Best PhD Award from EURASIP, the 2015 Ingvar Carlsson Award, and the 2014 Outstanding Young Researcher Award from IEEE ComSoc EMEA. He also co-authored papers that received best paper awards at the conferences WCSP 2017, IEEE ICC 2015, IEEE WCNC 2014, IEEE SAM 2014, IEEE CAMSAP 2011, and WCSP 2009.
\end{IEEEbiography}
\end{document}